\documentclass{ieeeaccess}
\usepackage{cite}
\usepackage{amsmath,amssymb,amsfonts}
\usepackage{algorithmic}
\usepackage{graphicx}
\usepackage{textcomp}
\usepackage{soul}
\usepackage{booktabs}%
\usepackage{bm}
\makeatletter
\AtBeginDocument{\DeclareMathVersion{bold}
\SetSymbolFont{operators}{bold}{T1}{times}{b}{n}
\SetSymbolFont{NewLetters}{bold}{T1}{times}{b}{it}
\SetMathAlphabet{\mathrm}{bold}{T1}{times}{b}{n}
\SetMathAlphabet{\mathit}{bold}{T1}{times}{b}{it}
\SetMathAlphabet{\mathbf}{bold}{T1}{times}{b}{n}
\SetMathAlphabet{\mathtt}{bold}{OT1}{pcr}{b}{n}
\SetSymbolFont{symbols}{bold}{OMS}{cmsy}{b}{n}
\renewcommand\boldmath{\@nomath\boldmath\mathversion{bold}}}
\makeatother

\def\BibTeX{{\rm B\kern-.05em{\sc i\kern-.025em b}\kern-.08em
    T\kern-.1667em\lower.7ex\hbox{E}\kern-.125emX}}

\begin{document}
\history{Received 27 March 2025, accepted 15 April 2025, date of publication 17 April 2025, date of current version 25 April 2025.}
\doi{10.1109/ACCESS.2025.3561857}

\title{DPN-GAN: Inducing Periodic Activations in Generative Adversarial Networks for High-Fidelity Audio Synthesis}
\author{\uppercase{Zeeshan Ahmad}\authorrefmark{1}, \IEEEmembership{Member, IEEE},
\uppercase{Shudi Bao}\authorrefmark{1}, \IEEEmembership{Member, IEEE},
AND \uppercase{Meng Chen} \authorrefmark{2}}

\address[1]{Ningbo Institute of Digital Twin, Eastern Institute of Technology, Ningbo 315200, People’s Republic of China}
\address[2]{School of Cyber Science and Engineering, Ningbo University of Technology, Ningbo 315211, People's Republic of China}

\tfootnote{This work was supported in part by the Open Research Fund of National Mobile Communications Research Laboratory, Southeast University (No.2023D15), Ningbo ClinicaI Research Center for Medical Imaging (No.2022LYKFYB01), and the Natural Science Foundation of Zhejiang, China (LGF22H120009).}

\markboth
{Z. Ahmad \headeretal: DPN-GAN: Inducing Periodic Activations in Generative Adversarial Networks for High-Fidelity Audio Synthesis}
{Z. Ahmad \headeretal: DPN-GAN: Inducing Periodic Activations in Generative Adversarial Networks for High-Fidelity Audio Synthesis}

\corresp{Corresponding author: Shudi Bao (e-mail: sdbao@idt.eitech.edu.cn).}

\begin{abstract}
In recent years, generative adversarial networks (GANs) have made significant progress in generating audio sequences. However, these models typically rely on bandwidth-limited mel-spectrograms, which constrain the resolution of generated audio sequences, and lead to mode collapse during conditional generation. To address this issue, we propose Deformable Periodic Network based GAN (DPN-GAN), a novel GAN architecture that incorporates a kernel-based periodic ReLU activation function to induce periodic bias in audio generation. This innovative approach enhances the model’s ability to capture and reproduce intricate audio patterns. In particular, our proposed model features a DPN module for multi-resolution generation utilizing deformable convolution operations, allowing for adaptive receptive fields that improve the quality and fidelity of the synthetic audio. Additionally, we enhance the discriminator network using deformable convolution to better distinguish between real and generated samples, further refining the audio quality. We trained two versions of the model: DPN-GAN small (38.67 M parameters) and DPN-GAN large (124M parameters). For evaluation, we use five different datasets, covering both speech synthesis and music generation tasks, to demonstrate the efficiency of the DPN-GAN. The experimental results demonstrate that DPN-GAN delivers superior performance on both out-of-distribution and noisy data, showcasing its robustness and adaptability. Trained across various datasets, DPN-GAN outperforms state-of-the-art GAN architectures on standard evaluation metrics, and exhibits increased robustness in synthesized audio.
\end{abstract}


\titlepgskip=-21pt

\maketitle

\section{Introduction}\label{sec:1}
\PARstart{A}{s} an emerging technique, generative adversarial networks (GANs) have been widely applied into various generation tasks, such as image generation, audio and speech synthesis, text generation, image translation, video generation, style-transfer, and so on \cite{cite_keyI1}. The GANs architecture includes two competing flexible networks: one is the generator that replicates a data distribution and generates synthesized data, the other is the discriminator which distinguishes between real and generated samples \cite{cite_keyI2}. The two opposing networks are trained alternately in a zero-sum game until a Nash equilibrium is reached. While GANs have demonstrated remarkable success in generating realistic and high-resolution images, they strive to achieve significant results in other domains \cite{cite_keyI1, cite_keyI2, cite_keyI3}. 

In recent years, the demand for high-quality and high-resolution audio has experienced explosive growth, driven by cutting-edge audio-related applications \cite{cite_keyI4}. For instance, speech-to-speech translation \cite{cite_keyI5} has enabled real-time translation of spoken languages, breaking down language barriers and facilitating global communication. Text-to-speech systems capable of handling numerous speakers \cite{cite_keyI6} have improved accessibility and personalized user experiences. Voice conversion technologies \cite{cite_keyI7} allow for the transformation of one speaker's voice to another, with applications in entertainment and privacy. Music generation \cite{cite_keyI8, cite_keyI9} has also benefited from advances in audio modeling, enabling the creation of original music compositions and soundscapes. Meanwhile, the frequent human-machine interactions also add up to the increased demand for the synthesis of high-quality human-like speech \cite{cite_keyI10}. However, modeling raw audio data presents significant challenges due to its high temporal resolution, which usually involves at least 16,000 samples per second \cite{cite_keyI11}. Additionally, audio data exhibits complex structures across different timescales, with dependencies ranging from short-term phonetic nuances to long-term prosodic patterns. These complexities make it difficult to accurately model and synthesize human-like speech. 

Recent approaches have focused on predicting low-resolution intermediate representations, such as mel-spectrograms, which capture the essential frequency components of audio signals \cite{cite_keyI12}. These intermediate representations are then used to synthesize raw waveform audio, leveraging the high-level features extracted from the mel-spectrograms. This approach has been employed in various state-of-the-art models, including WaveNet \cite{cite_keyI13} and Parallel WaveGAN \cite{cite_keyI12}, which have demonstrated the ability to produce high-fidelity audio. One of the key advancements in this area is the development of GAN-based vocoders, such as HiFi-GAN \cite{cite_keyI14}. These models can generate high-quality raw audio conditioned on mel-spectrograms, achieving synthesis speeds that are hundreds of times faster than on a single GPU in real time. This approach makes it feasible to deploy these models in real-world applications where low latency and high fidelity are crucial. Nevertheless, existing GANs models face severe challenges in modelling raw audio waveforms. They require a reasonable number of voices or speech data recorded in noise-free environment to generate high-resolution audio data. In addition, the quality of generated audios severely degrades when the model is conditioned on mel-spectrograms from unseen speakers in different acoustic environments \cite{cite_keyI15}. GANs also struggle with training difficulties and mode collapse, in which case the generator can only produces limited sample varieties \cite{cite_keyI1}. Denoising diffusion probabilistic models (DDPMs) have been proposed to address these issues but suffer from a slow reverse process, making them impractical for real-time applications. Additionally, existing GANs also struggle with balancing resolution and complexity. Increasing the resolution of the generated audio signals is computationally expensive and complex. While image generation models like super-resolution GAN \cite{cite_keyI16} have managed this challenge in the image domain, similar approaches for audio generation remain underexplored.

In this work, we propose DPN-GAN, a deformable periodic network-based GAN designed for the conditional generation of audio sequences with flexible resolution without the need for fine-tuning. The main contributions of this paper can be summarized as follows:
\begin{enumerate}
\item We propose a kernel-influenced ReLU-based periodic activation function, which allows us to control the resolution of the generated audio sequences by utilizing higher dimensional kernels for high-resolution generation, and lower dimensional activations for smaller sequences. 
\item We implement the DPN framework in the generator architecture for modeling complex audio waveforms. This framework employs a series of deformable convolution operations \cite{cite_keyI17} on the implicit representation of audio data, enabling the network to learn irregular patterns by using the adaptive kernel structures of deformable convolution \cite{cite_keyI18}. The network includes multi-resolution generating components with learnable periodicities, utilizing low-pass filters to reduce high-frequency signals, and high-pass filters to convert low-frequency signals into high-frequency responses. 
\item We propose a novel discriminator architecture inspired by the HiFi-GAN discriminator. Our discriminator architecture utilizes residual blocks consisting of deformable convolution, Position Sensitive Region of Interest Pooling (PSROIPooling), layer normalization, and periodic activation functions. This combination provides rich, and time-variant feature sets followed by non-linear transformations for effective classification. 
\item We demonstrate that DPN-GAN base with 38.67M parameters outperforms state-of-the-art audio generating networks in both in-distribution and out-of-distribution scenarios. Additionally, our DPN-GAN large (124M parameters) significantly surpasses existing models across various scenarios, including unseen speakers and recording environments. 
\end{enumerate}

The rest of this paper proceeds as follows. Section \ref{sec:2} reviews the related work about audio synthesis. Section \ref{sec:3} introduces preliminaries required for the proposed DPN-GAN. In section \ref{sec:4}, we introduce the architecture and working principle of the proposed DPN-GAN, followed by section \ref{sec:5} where we present the experimental setup to validate the performance of the proposed DPN-GAN. In section \ref{sec:6}, we conduct extensive experiments to evaluate the performance of the proposed DPN-GAN. Finally, section \ref{sec:7} concludes this paper. 

\textit{Notations:}
$|\cdot|$ , $\|\cdot\|$, and $\|\cdot\|_1$ are the modulus, Euclidean norm, and L1 distance, respectively. $\lfloor \cdot \rfloor$ means the floor function. $\Delta $ represents the adaptive shift of triangle waves in the AdaPReLU activation function. $\mathcal{R}$, $x$, and $y$ denote the regular grid, input feature map and output feature map, respectively. $\mathbb{E}(\cdot)$ stands for the expectation. $K$, $G(\cdot,\cdot)$ and $H$ denote the Gaussian kernel, interpolation kernel, and transfer function of a filter, respectively.

%%%%%%%%%%%%%%%%%%%%%%%%%%%%%%%%%%%%%%%%%%%%%%%%%%%%%%%%%%%%%%%%%%%%%%%
%%%%%%%%%%%%%%%%%%%%%%%%%%%%%%  SECTION 2 %%%%%%%%%%%%%%%%%%%%%%%%%%%%%%%%%%
%%%%%%%%%%%%%%%%%%%%%%%%%%%%%%%%%%%%%%%%%%%%%%%%%%%%%%%%%%%%%%%%%%%%%%%
\section{Related Work}\label{sec:2}
Existing studies on audio/speech synthesis can be broadly divided into four categories, including the pure signal processing techniques, autoregressive based models, non-autoregressive models, and GANs based models. In the following section, we will review the related works in detail.

%%%%%%%%%%%%%%%%%%%%%%%%%%%%%%%%%%
%%%%%%%%%%%%  Subsection 2.1 %%%%%%%%%%%%%%
%%%%%%%%%%%%%%%%%%%%%%%%%%%%%%%%%%

\subsection{Signal Processing Techniques}\label{subsec:2.1}
Griffin et al., \cite{cite_keyR1} proposed the Short-Time Fourier Transform (STFT) algorithm, which decodes an STFT sequence back to a temporal signal with noticeable artifacts. The algorithm is based on theoretical principles to estimate a signal from the modified STFT or its magnitude, and it is applied to problems like time-scale modification. Following this work, Wang et al., \cite{cite_keyR2} proposed the Tacotron model. It is an end-to-end generative Text-to-Speech (TTS) model, which uses a sequence-to-sequence framework with attention mechanism for converting text into raw spectrograms. With the emergence of Tacotron, TTS based models gains popularity in the signal processing community. Consequently, DeepVoice by Arik et al., \cite{cite_keyR3} optimizes the inference process to achieve real-time audio generation, simplifying TTS system creation. Then, DeepVoice 3 \cite{cite_keyR4} introduces a fully-convolutional architecture for speech synthesis, enhancing the training speed and scalability. A novel speech synthesis system, namely WORLD \cite{cite_keyR5}, employs a reliable F0 estimation algorithm, called distributed inline-filter operation, which improves the accuracy of pitch detection in speech signals. Sotelo et al, proposed Char2Wav, an end-to-end speech synthesis model that utilizes an attention-based bidirectional RNN encoder and a conditional Sample-RNN to map vocoder features to audio samples \cite{cite_keyR6}. More recently, Shen et al. \cite{cite_keyR7} proposed conditioning WaveNet model, which integrates a Tacotron-style model with a modified WaveNet vocoder to achieve high-quality speech synthesis. Furthermore, Ping et al. \cite{cite_keyR8} came up with Clarinet, which shows that a single variance-bounded Gaussian can model the raw waveform in WaveNet without compromising audio quality. However, it is difficult to accurately map intermediate features to audio in the aforementioned models, leading to obvious artifacts in the generated audio.

%%%%%%%%%%%%%%%%%%%%%%%%%%%%%%%%%%
%%%%%%%%%%%%  Subsection 2.2 %%%%%%%%%%%%%%
%%%%%%%%%%%%%%%%%%%%%%%%%%%%%%%%%%

\subsection{Autoregressive based models}\label{subsec:2.2}
Autoregressive based speech synthesis models can generate highly natural-sounding human speech, owing to their ability to capture long-term sequential dependencies in audio waveforms.  Oord et al. \cite{cite_keyR9} proposed a fully convolutional autoregressive model, called WaveNet, to generate high fidelity speech samples. It employs dilated casual convolutions to tackle the limitations of long-range dependencies in raw audio. Since their work on WaveNet, there has been substantial progress in the audio generation domains using autoregressive models. Following Oord's work, Mehri et al. \cite{cite_keyR10} proposed the SampleRNN model, a multiscale recurrent neural network (RNN) architecture that models raw audio at different temporal resolutions resulting in memory efficient training. Kalchbrenner et al. \cite{cite_keyR11} introduced a single layer RNN with a dual softmax layer, called WaveRNN. The quality of the output audio matches that of the WaveNet while being computationally efficient. WaveNet Autoencoders by Engel et al. \cite{cite_keyR12} captures long-term structure in audio through temporal hidden codes, contributing to the creation of the NSynth dataset for musical note synthesis. However, autoregressive based speech synthesis models suffer from slow inference speed due to inefficient sequential generation mechanism, and are therefore not feasible for real-world smart applications.

%%%%%%%%%%%%%%%%%%%%%%%%%%%%%%%%%%
%%%%%%%%%%%%  Subsection 2.3 %%%%%%%%%%%%%%
%%%%%%%%%%%%%%%%%%%%%%%%%%%%%%%%%%

\subsection{Non-autoregressive models} \label{subsec:2.3}
To tackle the issue of slow inference speed associated with autoregressive based models, non-autoregressive models were explored to parallelize the generation process. Oord et al. \cite{cite_keyI12} proposed Parallel WaveNet that introduces inverse autoregressive flows to improve the synthesis efficiency. It is also coupled with a novel neural network based distillation method for parallel training of a feed forward network from a trained WaveNet. Subsequently, Prenger et al. \cite{cite_keyR13} introduces WaveGlow model to show that autoregressive flows are unnecessary for speech synthesis. It uses a flow-based network with affine coupling layers and pointwise convolutions. NICE model proposed by Dinh et al. \cite{cite_keyR14} proposed a non-linear transformation which is easy to invert and has a tractable Jacobian matrix. Following their work, Kingma et al. \cite{cite_keyR15} proposed the model GLOW which builds on NICE and RealNVP with pointwise invertible convolutions and LU decomposition to enhance flow-based generative models. Although these models can significantly increase the inference speed, the quality of synthesized speech samples is inferior to autoregressive based models.  

%%%%%%%%%%%%%%%%%%%%%%%%%%%%%%%%%%
%%%%%%%%%%%%  Subsection 2.4 %%%%%%%%%%%%%%
%%%%%%%%%%%%%%%%%%%%%%%%%%%%%%%%%%

\subsection{GANs for Audio Generation} \label{subsec:2.4}
Recently, GANs have led to an increasing interest in audio generation domain. Kumar et al. \cite{cite_keyI11} proposed their pioneering work, MelGAN, which is a non-autoregressive and fully convolutional architecture generating audio waveforms by using induced receptive fields, multi-scale discriminator and multi-period discriminator. Following their work, Binkowski et al. \cite{cite_keyR16} proposed GAN-TTS model for text conditional high fidelity speech synthesis. They employed a convolutional generator and multiple discriminators evaluating different frequency ranges. Yamamoto et al. \cite{cite_keyR17} proposed parallel WaveGAN that optimizes WaveNet with multi-resolution STFT and adversarial loss functions to capture realistic speech waveforms. Subsequently Yang et al. \cite{cite_keyR18} demonstrated multiband MelGAN by increasing the receptive field of the MelGAN and using multi-resolution STFT loss, improving speech quality and training stability. The StyleMelGAN proposed by Mustafa et al. \cite{cite_keyR19} introduces a low-complexity GAN Vocoder with TADE layers, trained adversarially with multi-scale spectral reconstruction loss. Kong et al. \cite{cite_keyI14} presented their pioneering work on HiFi-GAN, employing a multi-receptive field fusion module and periodic discriminators to improve audio generation quality. Subsequently, Jang et al. \cite{cite_keyR20} proposed UnivNet, a real time neural vocoder with a multi-resolution spectrogram discriminator. Following HiFi-GAN, Morrison et al. \cite{cite_keyR21} introduced CarGAN which improves pitch accuracy in HiFi-GAN using an autoregressive conditioning stack. Efficient VAE by Lam et al. \cite{cite_keyR22} introduces MeLoDy, a diffusion model with dual-path diffusion and effective sampling schemes, and a successful audio VAE-GAN. PJLoop-GAN \cite{cite_keyR23} incorporates Projected GAN for audio-domain loop generation. Kim et al. \cite{cite_keyR24} proposed Fre-GAN that synthesizes frequency-consistent audio using resolution-connected generators and discriminators. Fre-GAN 2 by Lee et al. \cite{cite_keyR25} enhances Fre-GAN with fast and efficient synthesis using inverse discrete wavelet transform. VQCPC-GAN \cite{cite_keyR26} uses self-supervised training with Vector-Quantized Contrastive Predictive Coding to learn discrete representations for GAN-based audio generation. Liu et al. \cite{cite_keyR27} proposed UN-GAN which employs a hierarchical generator architecture and cycle regularization to avoid mode collapse in audio generation.

%%%%%%%%%%%%%%%%%%%%%%%%%%%%%%%%%%%%%%%%%%%%%%%%%%%%%%%%%%%%%%%%%%%%%%%
%%%%%%%%%%%%%%%%%%%%%%%%%%%%%%  SECTION 3 %%%%%%%%%%%%%%%%%%%%%%%%%%%%%%%%%%
%%%%%%%%%%%%%%%%%%%%%%%%%%%%%%%%%%%%%%%%%%%%%%%%%%%%%%%%%%%%%%%%%%%%%%%

\section{Preliminaries}\label{sec:3}
In this section, we will briefly review the deformable convolution operation and periodic ReLU activation function, which are relevant to the proposed DPN-GAN in section \ref{sec:4}.

%%%%%%%%%%%%%%%%%%%%%%%%%%%%%%%%%%
%%%%%%%%%%%%  Subsection 3.1 %%%%%%%%%%%%%%
%%%%%%%%%%%%%%%%%%%%%%%%%%%%%%%%%%
\subsection{One-Dimensional Deformable Convolution}\label{subsec:3.1}
The concept of deformable convolution for improving the geometric transformation modeling capability of CNNs was first proposed by Dai et al. \cite{cite_keyP1} in 2017. Originally, it was implemented with two-dimensional (2d) data. In this paper, however, we will be working with one-dimensional (1d) audio sequences datasets. Although the mathematical basis for deformable convolutions remains the same, the operation is modified for 1d domain. The idea of 1d deformable convolution is an extension of 2d deformable convolutions. The regular 1d convolution operation can be represented by following steps:
\begin{enumerate}
    \item Sampling using a regular grid $\mathcal{R}$ over the input space $\boldsymbol{x}$.
    \item Weighting the sampled values using a weight matrix $\boldsymbol{w}$.
    \item Summation of the weighted sample values.
\end{enumerate}

In a regular 1d convolutional network with a kernel size of $k$, the grid $\mathcal{R}$ can be defined as 
\begin{equation}
\label{eq1}
\mathcal{R} = \{ -\lfloor (\frac{k}{2}) \rfloor, \ldots, 0, \ldots, \lfloor (\frac{k}{2}) \rfloor \},
\end{equation}
where $\lfloor \cdot \rfloor$ denotes the floor function.

Each location $p_0$ on the output feature map $y$ can be defined as
\begin{equation}
\label{eq2}
y (p_0) = \sum_{p_n \in \mathcal{R}} w(p_n) \cdot x(p_0 + p_n).
\end{equation}

Deformable convolutions append offsets $\{ \Delta p_n | n = 1, \ldots, N \} $ to the grid $\mathcal{R}$, where $N = |\mathcal{R}|$. Thus, Eq. (\ref{eq2}) becomes
\begin{equation}
\label{eq3}
y(p_0) = \sum_{p_n \in \mathcal{R}} w (p_n) \cdot x (p_0 + p_n + \Delta p_n ).
\end{equation}

In general, the offset $\Delta p_n $ is fractional, therefore, sampling is performed via interpolation. For simplicity, we consider linear interpolation which is represented by Eq. \ref{eq4}.
\begin{equation}
\label{eq4}
    x(p) = \sum_q G(q, p) \cdot x(q),
\end{equation}
where $p$ denotes an arbitrary location, $q$ represents all the spatial locations in the feature map $x$, and $G(\cdot, \cdot)$ is the interpolation kernel. In 1d, the interpolation kernel $G$ is given by Eq. \ref{eq5}.
\begin{equation}
\label{eq5}
    G(q, p) = \max(0, 1 - |q - p|).
\end{equation}

The deformable convolutions in 1d domain involves two key processes: offset learning and bilinear interpolation. Offset learning is achieved by applying a convolutional layer over the same input feature map to obtain the offsets $\Delta p_{n}$. The convolution kernel used for generating these offsets has the same resolution and dilation as the one used in the deformable convolution, ensuring consistency in spatial alignment. As these offsets are often fractional, bilinear interpolation is employed to compute the values at these irregular locations. This interpolation method guarantees smooth transitions, and effective gradient propagation during backpropagation, which is crucial for the training process. Practically, deformable convolutions can adaptively adjust the receptive fields by learning these offsets during training. This adaptability allows for more flexible and informative feature extraction, which is particularly beneficial for tasks involving sequential data, such as audio signals or time-series analysis.

%%%%%%%%%%%%%%%%%%%%%%%%%%%%%%%%%%
%%%%%%%%%%%%  Subsection 3.2 %%%%%%%%%%%%%%
%%%%%%%%%%%%%%%%%%%%%%%%%%%%%%%%%%
\subsection{Position Sensitive RoI Pooling}\label{subsec:3.2}
PSRoIPooling is an advanced technique for handling RoIs in CNNs, particularly effective for object detection tasks. In PSRoIPooling, we first generate score maps from the input feature maps. These score maps are divided into $K$ bins. For each bin $(i,j)$, the pooled response for a given category $c$, $r_c (i, j| \Theta)$, is computed as
\begin{equation}
\label{eq6}
r_c(i, j|\Theta) = \sum_{(x, y) \in \text{bin}(i,j)} z_{i,j,c}(x + x_0, y + y_0|\Theta) / n,
\end{equation}
where $z_{i,j,c}$ is the score map for category $c$, $(x_0, y_0)$ denotes the top-left corner of the RoI, $n$ denotes the number of pixels in the bin, and the parameter $\Theta$ represents all learnable parameters of the network.
For 1d data, the RoI is specified by its starting point $x_0$ and length $l$. The pooled response for each bin $i$ is then calculated as
\begin{equation} 
\label{eq7}
r_c(i|\Theta) = \sum_{x \in \text{bin}(i)} z_{i,c}(x + x_0|\Theta) / n.
\end{equation}

In this way, the PSRoI pooling operation captures fine-grained positional information in each RoI. Moreover, it provides a structured approach to handle varying lengths of input sequences, ensuring that the positional context is utilized effectively during the feature extraction process.

%%%%%%%%%%%%%%%%%%%%%%%%%%%%%%%%%%%%%%%%%%%%%%%%%%%%%%%%%%%%%%%%%%%%%%%
%%%%%%%%%%%%%%%%%%%%%%%%%%%%%%  SECTION 4 %%%%%%%%%%%%%%%%%%%%%%%%%%%%%%%%%%
%%%%%%%%%%%%%%%%%%%%%%%%%%%%%%%%%%%%%%%%%%%%%%%%%%%%%%%%%%%%%%%%%%%%%%%
\section{Proposed Method}\label{sec:4}
In this section, we propose DPN-GAN to synthesize high-fidelity diverse audio samples with flexible resolution. The first subsection designs an Adaptive Periodic ReLU (AdaPReLU) activation function used in the proposed model. The next subsection presents the architectural descriptions of the generator and discriminator networks. The final subsection provides the loss functions used to train the proposed DPN-GAN.

%%%%%%%% Figure 1 %%%%%%%%%%%%%
\begin{figure*}[htbp]
    \centering
    \includegraphics[width=1\linewidth]{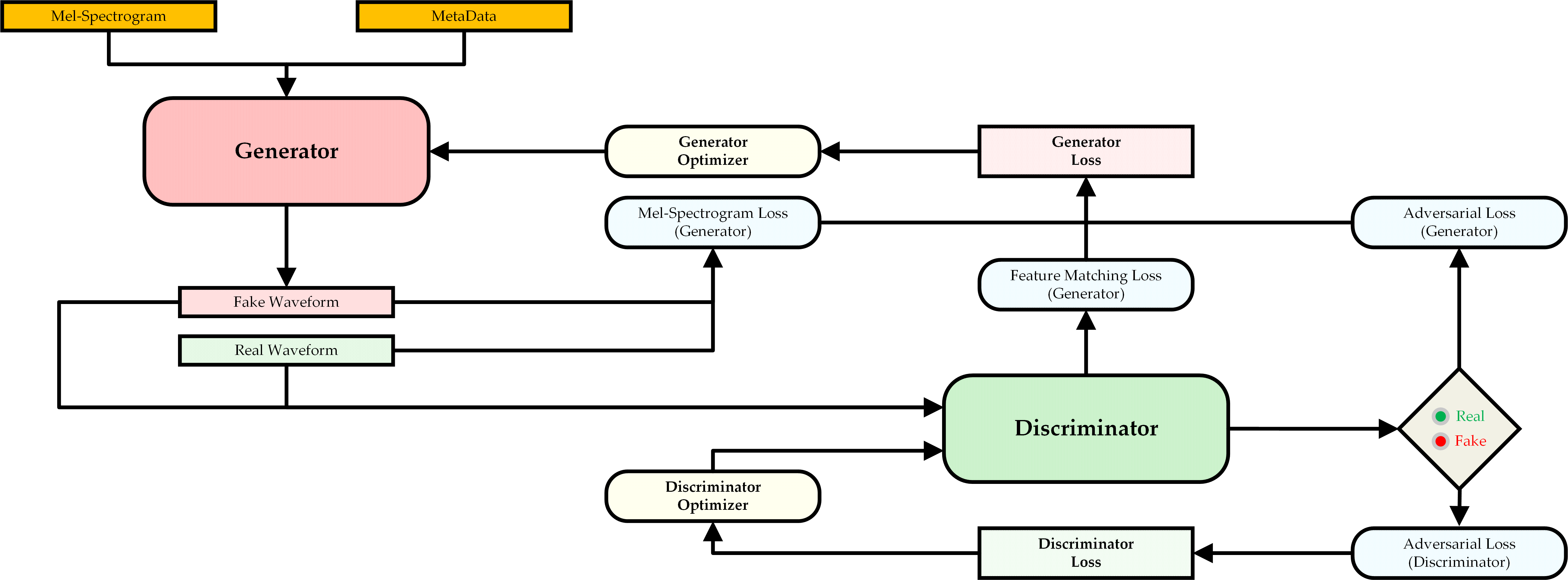}
    \caption{Flowchart of our proposed DPN-GAN}
    \label{fig1}
\end{figure*}
%%%%%%%%%%%%%%%%%%%%%%%%%%

%%%%%%%%%%%%%%%%%%%%%%%%%%%%%%%%%%
%%%%%%%%%%%%  Subsection 4.1 %%%%%%%%%%%%%%
%%%%%%%%%%%%%%%%%%%%%%%%%%%%%%%%%%
\subsection{Adaptive Periodic ReLU} \label{subsec:4.1}
We propose an adaptive periodic activation function, which is inspired by the work of Meronen et al. \cite{cite_keyD1}. They proposed a periodic ReLU activation function by summing up two triangle waves, with the second one being shifted by one-fourth of a period. Through analysis, we observed that the periodic nature of the wave has variance for different audio signals. So, we choose an adaptive parameter to modify the shift of the second triangle wave.

The periodic ReLU activation function can be expressed as
\begin{multline}
\label{eq8}
\psi_{PReLU}(x) = \frac{8}{\pi^2} \left(\left( (x + \frac{\pi}{2}) - \pi \left\lfloor \frac{(x + \frac{\pi}{2})}{\pi} + \frac{1}{2} \right\rfloor \right) \right.\\
\left. (-1)^{ \lfloor \frac{(x + \frac{\pi}{2})}{\pi} + \frac{1}{2} \rfloor} + \left( x - \pi \left\lfloor \frac{x}{\pi} + \frac{1}{2} \right\rfloor \right) (-1)^{ \lfloor \frac{x}{\pi} + \frac{1}{2} \rfloor }\right),
\end{multline}
where $p=2\pi$ is the considered period. To introduce adaptivity in the shifts of the triangle waves, we let the shifts to be learnable parameters. Let $\Delta$ represent the adaptive shift, the AdaPReLU activation function is given by Eq. \ref{eq9}.
\begin{multline}
\label{eq9}
\psi_{AdaPReLU}(x) = \frac{8}{\pi^2} \left( \!\! \left(\!\! (x + \Delta) - \pi \left\lfloor \frac{(x + \Delta)}{\pi} + \frac{1}{2} \right\rfloor \right) \right.\\
\left.  (-1)^{ \lfloor \frac{(x + \Delta)}{\pi} + \frac{1}{2} \rfloor }  + \left( (x - \Delta) - \pi \left\lfloor \frac{(x - \Delta)}{\pi} + \frac{1}{2} \right\rfloor \right) \right.\\
 \left. (-1)^{ \lfloor \frac{(x - \Delta)}{\pi} + \frac{1}{2}\rfloor } \right).
\end{multline}

We allow $\Delta$ to be learned during the training process. The AdaPReLU activation function can more effectively capture the underlying patterns in the data, providing a flexible and powerful activation mechanism. This adaptivity enhances the model's ability to approximate complex functions and improves its performance on tasks involving non-stationary signals.

%%%%%%%%%%%%%%%%%%%%%%%%%%%%%%%%%%
%%%%%%%%%%%%  Subsection 4.2 %%%%%%%%%%%%%%
%%%%%%%%%%%%%%%%%%%%%%%%%%%%%%%%%%
\subsection{Architecture of DPN-GAN}\label{subsec:4.2}
Fig. \ref{fig1} illustrates an overview of the proposed DPN-GAN for generating high fidelity diverse audio signals of flexible resolution. It takes mel-spectrogram and metadata of the audio as input to the generator network, and outputs a generated audio signal. Following this, both the generated signal and real signal passes through the discriminator network which classifies each signal accordingly. Then, we compute various losses associated with the generator and discriminator networks. The generator loss consists of three different loss components, which are adversarial loss for the generator, mel-spectrogram loss of the generated and real audio signals, and feature matching loss between the generated audio signal and real audio signal passing through the discriminator. On the other hand, the discriminator connects to the adversarial loss only. In the following subsections, we will discuss each component of the DPN-GAN in detail.

%%%%%%%%% Subsubsection 4.2.1 %%%%%%%%%%%
\subsubsection{Generator}\label{subsubsec:4.2.1}

\begin{figure*}[htbp]
    \centering
    \includegraphics[width=0.8\linewidth]{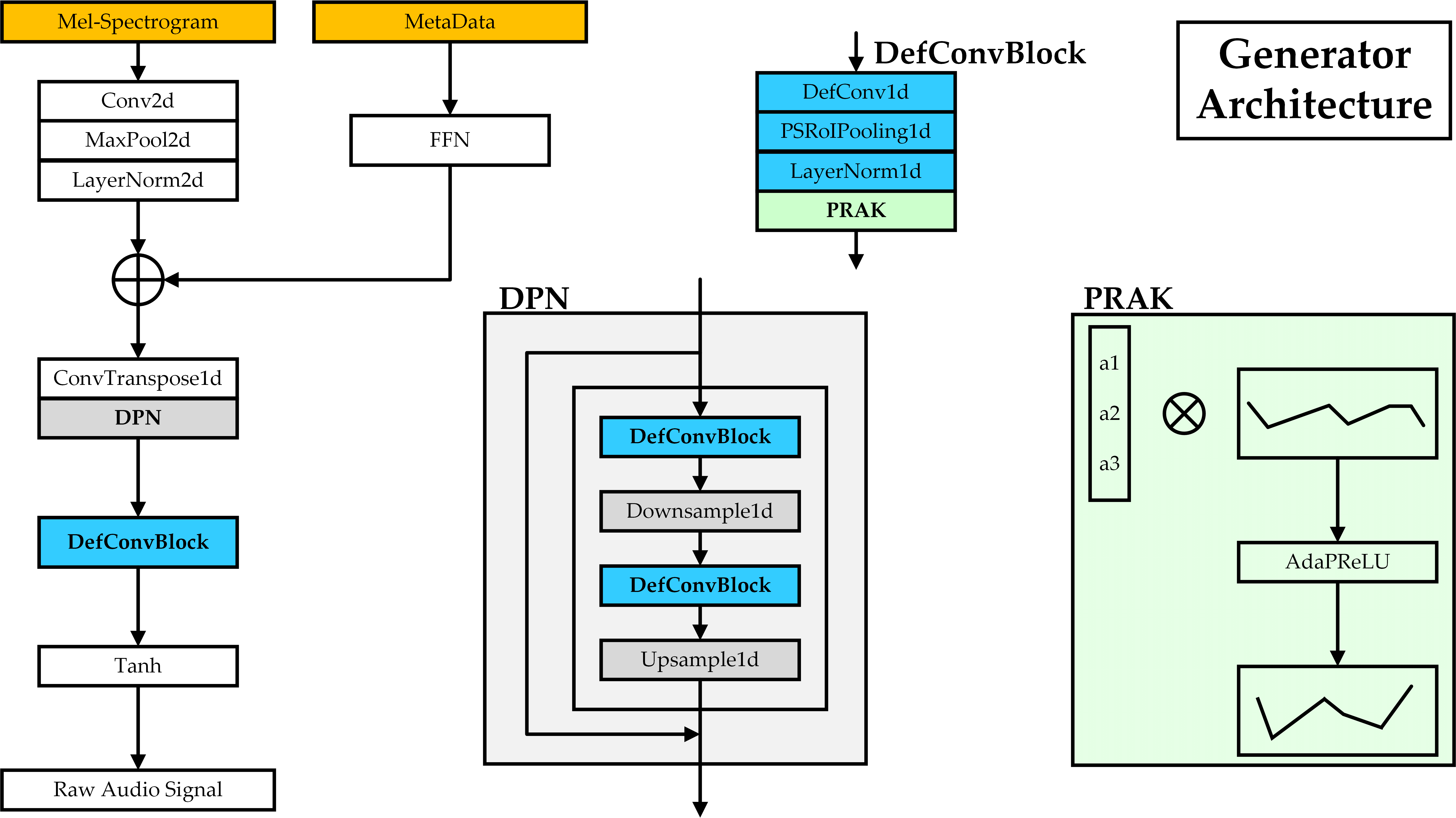}
    \caption{Generator architecture of DPN-GAN}
    \label{fig2}
\end{figure*}

The architecture of the proposed generator network is shown in Fig. \ref{fig2}. It is a fully convolutional network based model, which takes the mel-spectrogram and metadata as input, and generates raw audio waveforms. It should be noted that mel-spectrogram is 2d in nature, whereas metadata is a 1d vector. First, the mel-spectrogram is passed through a convolutional block, which applies convolution operation followed by max-pooling and layer normalization. This extracts initial features from the data, and layer normalization provides consistency to the feature space. On the other hand, metadata is passed through a series of fully connected layers to extract useful information from it. Subsequently, these two extracted features are concatenated, and the combined output is then passed to the following layers. Next, we increase the resolution of the extracted features by passing them through a 1d transpose convolution layer. After upscaling the features, it passes through the DPN layer.

%%%%%%%% Figure 2 %%%%%%%%%%%%%
%%%%%%%%%%%%%%%%%%%%%%%%%%

\textbf{Deformable Periodic Network:} We have designed a DPN module for our generator to observe extracted features at different scales. It is a combination of multilayer residual blocks, where the output of each layer gets summed up, and provides the output of the DPN module. Inside each residual layer, the input passes through a deformable convolution block (DefConvBlock). The structure of the DefConvBlock is shown in the architecture diagram of the generator (Fig. \ref{fig2}). In the DefConvBlock, the input passes through a deformable convolution layer followed by a PSRoIPooling layer. After pooling operation, it passes through a layer normalization operation, and an activation layer of AdaPReLU weighted by a kernel matrix. This activation block is called as Periodic ReLU Activated Kernel (PRAK).

\textbf{Periodic ReLU Activated Kernel:} We have implemented a specific activation kernel for modelling purpose. In section \ref{subsec:4.1}, we proposed an AdapReLU activation function, which enhances the model's ability to approximate complex functions, and improves its performance on tasks involving non-stationary signals. In PRAK block, we first integrate the feature space with the Gaussian kernel to bring the space to a normal distribution. The Gaussian kernel is defined by the function:
\begin{equation}
\label{eq10 }
K(x, x') = \exp\left(-\frac{\|x - x'\|^2}{2\sigma^2}\right),
\end{equation}
where $x$ and $x'$ are input vectors, $\|x-x’\|$ represents the Euclidean distance between these vectors, and $\sigma$ is the bandwidth parameter that controls the width of the Gaussian. The Gaussian kernel is characterized by smooth bell-shaped curve, which ensures that points closer in the input space have higher similarity. This property makes the Gaussian kernel effective in capturing local structures and smoothing noisy data. The smoothness parameter $\sigma$ of the kernel plays a crucial role in determining the extent of the smoothing, with smaller values leading to narrower and more localized effects, whereas larger values resulting in broader and more global smoothing.

Following this, we apply the AdaPReLU activation function to this normalized feature space to obtain the non-linear periodic output. Next, we downsample the output of the DefConvBlock using a low pass filter. The transfer function of an ideal low pass filter is given by

\begin{equation}
\label{eq11}
    H_{\text{LP}}(\omega) = 
    \begin{cases} 
    1 & \text{if } |\omega| \leq \omega_c \\
    0 & \text{if } |\omega| > \omega_c
    \end{cases}
\end{equation}
where $\omega$ and $\omega_c$ denote the angular frequency and cutoff frequency, respectively. However, since the ideal transfer function is not differentiable in nature, we apply the algebraic function of the low pass filter given by
\begin{equation}
\label{eq12}
    H_{\text{LP}}(\omega) = \frac{1}{1 + j\frac{\omega}{\omega_c}},
\end{equation}

After reducing the scale of the feature space, we again pass it through a DefConvBlock followed by an upsampling layer using a high pass filter. The transfer function of an ideal high-pass filter is given by
\begin{equation}
\label{eq13}
    H_{\text{HP}}(\omega) = 
    \begin{cases} 
    0 & \text{if } |\omega| \leq \omega_c \\
    1 & \text{if } |\omega| > \omega_c 
    \end{cases}
\end{equation}

Again, during implementation of the model, we apply the algebraic form of high pass filter which is given by
\begin{equation}
\label{eq14}
    H_{\text{HP}}(\omega) = \frac{j\frac{\omega}{\omega_c}}{1 + j\frac{\omega}{\omega_c}},
\end{equation}

By using these downsampling and upsampling layers, the model learns features at different resolutions. These extracted features are further processed using a DefConvBlock and a Tanh activation function. The final output represents the raw generated audio signal.

%%%%%%%%% Subsubsection 4.2.2 %%%%%%%%%%%
\subsubsection{Discriminator}\label{subsubsec:4.2.2}

%%%%%%%% Figure 3 %%%%%%%%%%%%%
\begin{figure*}[htbp]
    \centering
    \includegraphics[width=0.8\linewidth]{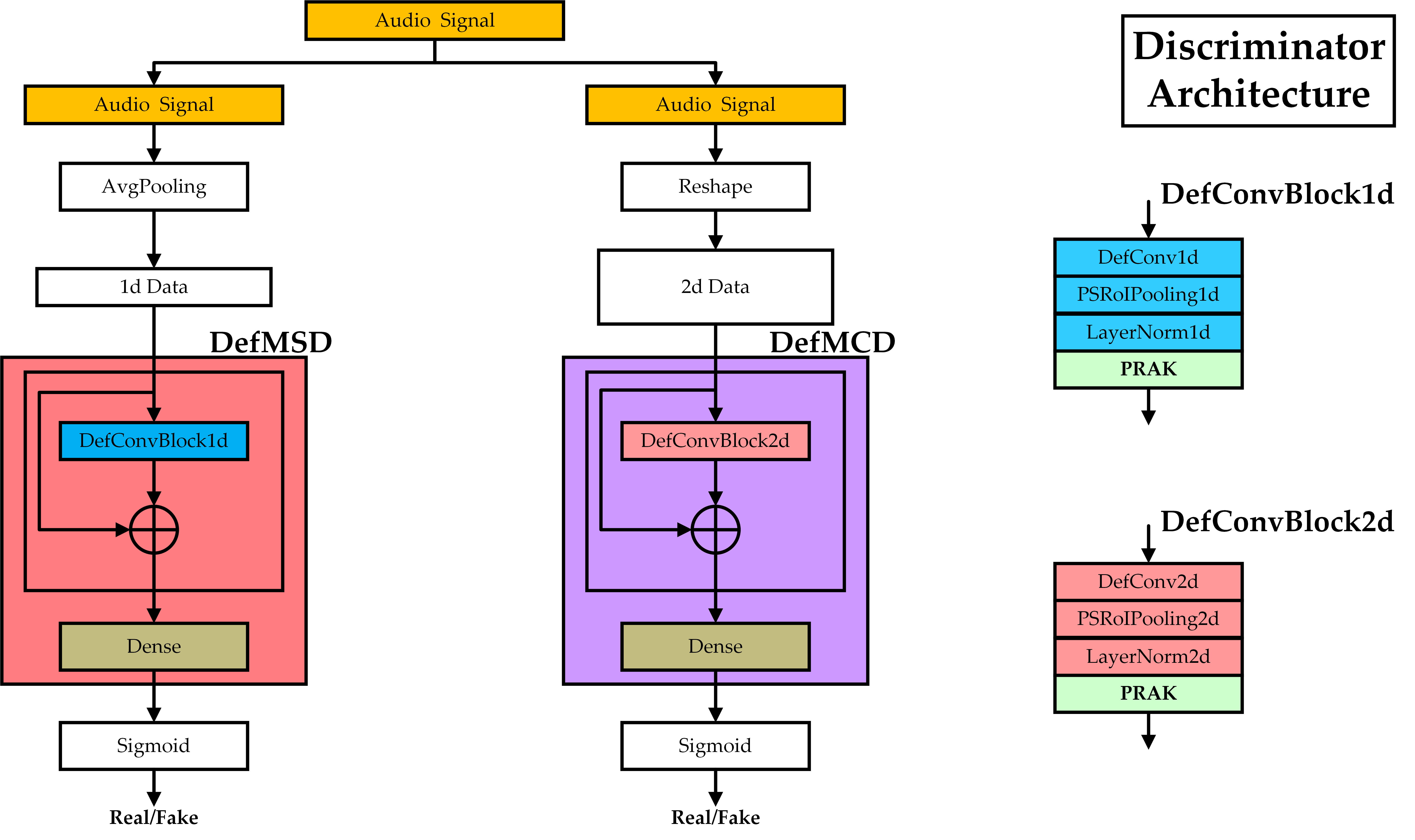}
    \caption{Discriminator architecture of DPN-GAN}
    \label{fig3}
\end{figure*}
%%%%%%%%%%%%%%%%%%%%%%%%%%
The structure of our discriminator network is inspired by the configuration of HiFi-GAN discriminator \cite{cite_keyI14}. Fig. \ref{fig3} shows the architecture of our discriminator network. It has two different modules: the Deformable Multi-Scale Discriminator (DefMSD) and the Deformable Multi-Channnel Discriminator (DefMCD). The former extracts features from the input space at different resolutions, the latter process the periodic samples inside the audio signal. First, we create two different feature spaces from the input audio signal. For input to DefMSD, we reduce the size of the audio signal using average pooling. On the other hand, we reshape the 1d audio signal into 2d data for channel wise processing in DefMCD. 

\textbf{Deformable Multi Channel Discriminator:}
DefMCD is a mixture of sub-discriminators, which takes the audio input at equally spaced intervals. The audio signal is converted into 2d feature space to consider the periodic nature of the audio signal. The period $p$ is a user-defined parameter set to be equally spaced interval length. The sub-discriminators have same architecture, i.e, a series of 2d DefConv blocks, but the inputs they take varies due to considered periodicity. Each DefConvBlock stacks a number of operations, including deformable convolution, PSRoIPooling, layer normalization and PRAK activation kernel. For the deformable convolution operation, we consider different kernel sizes from [2, 3, 5, 7, 11] to extract features at different scales of the discretized audio input. Finally, the outputs from each kernel are concatenated for further operation. Since we have used residual blocks, the input signal is concatenated to the output signal after the DefConvBlock operation. Each sub-discriminator captures different implicit structures of the audio sequence by looking at different parts of the audio data. Therefore, we set different period values to avoid overlapping operations. We then pass the output of the residual deformable convolution block through a dense layer to obtain the feature representation of DefMCD. By transforming the input audio into 2D data rather than sampling its periodic signals, gradients from the DefMCD can be propagated to every time step of the input audio.

\textbf{Deformable Multi Scale Discriminator:}
DefMSD is also a mixture of sub-discriminators, which takes the reduced sequence of the audio input. The output of average pooling layer goes into the DefMSD sub-discriminators as input. In DefMCD, we convert each audio signal into discrete feature space on the basis of period. Here, we implement the DefMSD architecture to conserve the correlation between the discrete feature spaces. Each sub-discriminator in DefMSD consists of 1d DefConvBlock. This block consists of a 1d deformable strided convolution operation, 1d PSRoIPooling operation, 1d layer normalization, and a PRAK activation kernel. To process the lengthy input signal at different resolution, we consider various kernel sizes ranging from 2 to 11. The output of each sub-discriminator is concatenated to each other for further processing. Again, we use residual connections here. Hence, the output signal gets concatenated to the input signal. Following this, the outputs are passed through dense layers to obtain the final feature representation of DefMSD. One innate difference between the DefMSD and DefMCD is that, while DefMCD operates on discretized samples of raw waveform, DefMSD operates on smoothed waveforms.

The output of both DefMSD and DefMCD finally passes through a fully connected layers of two nodes and a sigmoid function which outputs the binary classification results as shown in the discriminator architecture.

%%%%%%%%%%%%%%%%%%%%%%%%%%%%%%%%%%
%%%%%%%%%%%%  Subsection 4.3 %%%%%%%%%%%%%%
%%%%%%%%%%%%%%%%%%%%%%%%%%%%%%%%%%
\subsection{Losses} \label{subsec:4.3}

For training, the loss function of our DPN-GAN follows the MelGAN \cite{cite_keyI11} and HiFi-GAN \cite{cite_keyI14}. In Fig. \ref{fig1}, we have shown different loss functions associated with the generator and discriminator networks of the DPN-GAN. Similar to HiFi-GAN, we employ the feature matching loss and mel-spectrogram loss in addition to the standard adversarial loss to train our generator network. On the other hand, the discriminator network considers only the adversarial loss for its training. Moreover, the standard adversarial loss is replaced by least-squares formulation functions for non-vanishing gradient flows \cite{cite_keyL1}, as in the MelGAN. 

The least-square adversarial loss, $\mathcal{L}_{\text{Adv}}(G;D)$, aims to fool the discriminator can be expressed as Eq. \ref{eq15}:
\begin{equation}
\label{eq15}
\mathcal{L}_{\text{Adv}}(G;D) = \mathbb{E}_s \left[ (D(G(s)) - 1)^2 \right],
\end{equation}
where $G(s)$ represents the generated sample from the input condition $s$, and $D$ is the discriminator. 

The feature matching loss, $\mathcal{L}_{\text{FM}}(G;D)$, helps to stabilize the training process by minimizing the L1 distance between the feature representations of real and generated samples across multiple layers of the discriminator. It is given by Eq. \ref{eq16}:
\begin{equation}
\label{eq16}
\mathcal{L}_{\text{FM}}(G;D) = \mathbb{E}_{(x,s)} \left[ \sum_{i=1}^{T} \frac{1}{N_i} \| D_i(x) - D_i(G(s)) \|_1 \right],
\end{equation}
where $D_i$ denotes the features and $N_i$ is the number of corresponding features in the $i$-th layer of the discriminator. 

The mel-spectrogram loss, $\mathcal{L}_{\text{Mel}}(G)$, measures the L1 distance between the mel-spectrograms of the generated and ground truth waveforms, enhancing the perceptual quality of the synthesized audio. It is defined by Eq. \ref{eq17}:
\begin{equation}
\label{eq17}
\mathcal{L}_{\text{Mel}}(G) = \mathbb{E}_{(x,s)} \left[ \| \phi(x) - \phi(G(s)) \|_1 \right],
\end{equation}
where $\phi$ is the function that maps a waveform to its respective mel-spectrogram. 

The final generator loss is the weighted sum of these three loss functions:
\begin{equation}
\label{eq18}
\mathcal{L}_G = \mathcal{L}_{\text{Adv}}(G;D) + \lambda_{\text{FM}} \mathcal{L}_{\text{FM}}(G;D) + \lambda_{\text{Mel}} \mathcal{L}_{\text{Mel}}(G),
\end{equation}
where $\lambda_{\text{FM}}$ and $\lambda_{\text{Mel}}$ denote the weights for the feature matching and mel-spectrogram losses, respectively.

The discriminator loss is solely based on the adversarial loss, which is designed to distinguish between real and generated samples. The discriminator is trained to classify the real audio samples as close to 1 and the generated samples as close to 0. The least-square adversarial loss, $\mathcal{L}_{\text{Adv}}(D;G)$, for the discriminator is formulated as Eq. \ref{eq19}:
\begin{equation}
\label{eq19}
\mathcal{L}_{\text{Adv}}(D;G) = \mathbb{E}_{(x,s)} \left[ (D(x) - 1)^2 + (D(G(s)))^2 \right],
\end{equation}
where $x$ denotes the ground truth audio. This loss ensures that the discriminator effectively learns to differentiate between real and synthetic audio, providing meaningful gradients to the generator for improving the realism of the generated samples.

These two objective losses (generator loss and discriminator loss) are optimized separately in a contrastive manner, which follows the training principle of GANs.

%%%%%%%%%%%%%%%%%%%%%%%%%%%%%%%%%%%%%%%%%%%%%%%%%%%%%%%%%%%%%%%%%%%%%%%
%%%%%%%%%%%%%%%%%%%%%%%%%%%%%%  SECTION 5 %%%%%%%%%%%%%%%%%%%%%%%%%%%%%%%%%%
%%%%%%%%%%%%%%%%%%%%%%%%%%%%%%%%%%%%%%%%%%%%%%%%%%%%%%%%%%%%%%%%%%%%%%%
\section{Experimental Setup}\label{sec:5}
We implement the proposed DPN-GAN with PyCharm 2022.2.1, and Google Colab with Python 3.9.13. All experiments were performed on a 10th generation intel i5 processing system with $16$ GB RAM and $8$ processing threads. Additionally, we used an 8 GB GPU computing resource for experiments.

%%%%%%%%%%%%%%%%%%%%%%%%%%%%%%%%%%
%%%%%%%%%%%%  Subsection 5.1 %%%%%%%%%%%%%%
%%%%%%%%%%%%%%%%%%%%%%%%%%%%%%%%%%

\subsection{Datasets}\label{subsec:5.1}

For training and evaluation, we used four benchmark speech datasets including the LJSpeech \cite{cite_keyE1}, VCTK \cite{cite_keyE2}, LibriSpeech \cite{cite_keyE3} and AudioMNIST \cite{cite_keyE4}, to test our proposed DPN-GAN model. Besides the above-mentioned datasets, to show that the proposed DPN-GAN can also be applied to non-speech (music) datasets, we also conduct the evaluations on the GTZAN dataset \cite{cite_keyE5}. The details of the datasets are presented as follows.

\textbf{The LJSpeech Dataset:} is a collection of 13,100 short audio clips accompanied by a transcription, featuring a single speaker reading passages from seven non-fiction books. The duration of each clip varies from 1 to 10 seconds, totaling approximately 24 hours of speech data. 

\textbf{VCTK Dataset:} is a multi-speaker corpus composed of speech data from 109 native English speakers with different accents. Each speaker reads about 400 sentences, including selections from newspapers, the rainbow passage, and an elicitation paragraph to identify the speaker's accent. This dataset is particularly useful for building speaker-adaptive text-to-speech synthesis systems.

\textbf{LibriSpeech Dataset:} is a large-scale corpus composed of nearly 1000 hours of 16kHz read English speech, derived from audiobooks that were part of the LibriVox project. The dataset is divided into different subsets including a "clean" subset that contains recordings with minimal background noise and higher audio quality, and an "other" subset with recordings in more challenging conditions including higher levels of noise and less clear speech. Each subset is further split into training, validation, and test sets. For our analysis, we have considered the split which contains 100 hours of recording. 

\textbf{AudioMNIST Dataset:} is comprised of 30,000 audio recordings of spoken digits (0–9) in English, with each digit repeated 50 times by 60 different speakers. Recorded in quiet offices using a RØDE NT-USB microphone at a sampling frequency of 48 kHz, the dataset totals about 9.5 hours of speech. Meta-information such as age, sex, origin, and accent of the speakers is also included. The dataset is used for benchmarking models for various classification tasks, including digit, speaker and sex classification.

\textbf{GTZAN Dataset:} is a well-known collection for music genre classification tasks. It consists of a total of 1000 audio files, equally divided across 10 distinct genres. Each genre contains 100 audio files, with each file having a duration of 30 seconds. The dataset includes mel-spectorgrams for each audio file, facilitating various audio analysis and visualization tasks. Additionally, the dataset provides feature extraction details, offering two types of files: one with mean and variance computed over multiple features for the entire 30-second audio files, and another with the same structure but calculated over 3-second segments, obtained by splitting the original 30-second files.

%%%%%%%%%%%%%%%%%%%%%%%%%%%%%%%%%%
%%%%%%%%%%%%  Subsection 5.2 %%%%%%%%%%%%%%
%%%%%%%%%%%%%%%%%%%%%%%%%%%%%%%%%%
\subsection{Model Configuration}\label{subsec:5.2}
We trained two versions of the model: DPN-GAN small and DPN-GAN large. For the data configuration of the AudioMNIST dataset, the number of input channels for the mel-spectrograms is $1$, the number of filters is $128$, and there are $100$ time frames in the dataset. 

In the generator network, we used $32$ hidden channels with a kernel size of $3$ for the mel-spectrogram initiator layer. The size of the metadata input structures are $(1, 152)$. The hidden dimension of the metadata initiator layer is set to $64$ for DPN-GAN small and $128$ for DPN-GAN large. Following these two layers, we have considered a $3 \times 3$ kernel in the DPN block with $64$ hidden channels for DPN-GAN small and $512$ hidden channels for DPN-GAN large. Finally, we convert the DPN output into the audio sequence using a dense layer with a hidden dimension size of $47749$. 

The discriminator network considered an average pooling kernel size and stride of $11$ and $4$, respectively, in the MSD preprocessing layer for both versions. In the sub-discriminator of the MSD module, we have taken a range of kernel sizes from $[3, 5, 7, 11]$ with a stride of $1$ in each sub-discriminator layer, and a final layer stride of $4$. Similarly, in the SubMCD layers, we have considered the kernel sizes of $[3, 5, 7, 11]$ for variable scale of feature processing. For both the layers, the hidden dimension of the final layer is set to $512$ for DPN-GAN small and $2048$ for DPN-GAN large, and the output dimension size of $2$ is considered for binary classification.

%%%%%%%%%%%%%%%%%%%%%%%%%%%%%%%%%%
%%%%%%%%%%%%  Subsection 5.3 %%%%%%%%%%%%%%
%%%%%%%%%%%%%%%%%%%%%%%%%%%%%%%%%%

\subsection{Performance Metrics}\label{subsec:5.3}
Since we used both speech and music datasets, five well-established metrics covering both application scenarios are chosen for evaluating the model performance. Perceptual Evaluation of Speech Quality (PESQ) \cite{cite_keyE6}, Short-Time Objective Intelligibility (STOI) \cite{cite_keyE7} and WARP-Q \cite{cite_keyE8} are metrics used to estimate the quality of speech generation, whereas, Fr\'echet Audio Distance (FAD) \cite{cite_keyE9}, Fr\'echet Deep Speech Distance (FDSD) \cite{cite_keyR16} evaluates the quality of generated music of the generative model. 

\textbf{PESQ:} is a widely used metric to assess the perceptual quality of synthesized speech samples. It computes the absolute difference between the degraded and a reference signal, which are pre-processed through several steps to extract distortions, and a non-linear average is calculated over time and frequency. It integrates the disturbance over several time scale, which is then aggregated using $L_{p}$ norm. The range of PESQ score is from -0.5 to 4.5, with higher score indicating better perceptual quality of the synthesized speech audio. 

\textbf{Short-Time Objective Intelligibility (STOI):} is a widely-used metric that predicts speech intelligibility, particularly in speech enhancement and noise reduction contexts. It computes the cross correlation between the temporal envelopes of a clean and degraded speech for short overlapping segments. The metric scores range from 0 to 1, where higher scores means that the speech signal is more intelligible and easier to understand.

\textbf{WARP-Q:} calculates an optimal matching cost between two given audio sequences. It is based on DTW (Dynamic Time Window), which is a well-known metric used in a number of speech processing applications. Unlike traditional speech quality metrics, WARP-Q handles the time-alignment and signal similarity in a combined manner. The distance between two speech signals is measured by using subsequence DTW (SDTW), which search for the numerous matches of a sequence within the longer sequence.

\textbf{Fr\'echet Audio Distance (FAD):} is a reference-free evaluation metric based on Fr\'echet Inception Distance (FID) designed for music enhancement models. It computes the Fr\'echet distance between the distribution of embedding statistics generated on the whole evaluation set and that on a reference set of clean music. 

\textbf{ Fr\'echet Deep Speech Distance (FDSD):} is a metric designed to assess the quality of generated speech against real speech samples. It adapts the FID concept, originally used in image processing, to the audio domain by utilizing deep speech representations. The FDSD score ranges from 0 to $\infty$, with lower FDSD value indicating higher similarity between the synthetic and real distributions.

%%%%%%%%%%%%%%%%%%%%%%%%%%%%%%%%%%%%%%%%%%%%%%%%%%%%%%%%%%%%%%%%%%%%%%%
%%%%%%%%%%%%%%%%%%%%%%%%%%%%%%  SECTION 6 %%%%%%%%%%%%%%%%%%%%%%%%%%%%%%%%%%
%%%%%%%%%%%%%%%%%%%%%%%%%%%%%%%%%%%%%%%%%%%%%%%%%%%%%%%%%%%%%%%%%%%%%%%
\section{Results and Discussion}\label{sec:6}
In this section, we conducted extensive experiments to demonstrate the performance of the proposed DPN-GAN. The first subsection evaluates the performance of the DPN-GAN on various datasets, and the results are compared with several state-of-the-art GAN models. The next subsection presents ablation experiments to illustrate the impact of key components and loss functions of the proposed DPN-GAN. The third subsection demonstrates the model's performance on out-of-distribution and noisy data. The final subsection provides the runtime comparison.

%%%%%%%%%%%%%%%%%%%%%%%%%%%%%%%%%%
%%%%%%%%%%%%  Subsection 6.1 %%%%%%%%%%%%%%
%%%%%%%%%%%%%%%%%%%%%%%%%%%%%%%%%%
\subsection{Performance on Different Datasets}\label{subsec:6.1}
In this subsection, we compare the performance of the DPN-GAN with other state-of-the-art generative models on AudioMNIST, LJSpeech, LibriSpeech, and VCTK datasets. In addition, we also analyze various training parameters such as training data size, depth of DPN layer, the depth of MSD and MCD on the AudioMNIST dataset. In these experiments, five methods are used for comparison, including HiFi-GAN \cite{cite_keyI14}, UNIV-NET \cite{cite_keyR20}, SpecDiff-GAN \cite{cite_keyE10}, BigVGAN \cite{cite_keyI15}, and Fre-GAN \cite{cite_keyR24}.

%%%%%%%%% Subsubsection 6.1.1 %%%%%%%%%%%
\subsubsection{Results on AudioMNIST} \label{subsubsec:6.1.1}

We first evaluate the effect of training data size on the performance of the proposed DPN-GAN in order to find optimum train-test split. The PESQ, STOI and WARP-Q scores for the proposed DPN-GAN with different sizes of the training dataset are tabulated in Table \ref{tab1}. From the Table \ref{tab1}, we can see that as the size of the training dataset increases, the performance of the proposed DPN-GAN gets improved. Considering $1\%$ of the data for training, we observe that the performance metrics are poor. Generally, GANs require substantial amount of data to train effectively. By increasing the training data size to $90\%$, the proposed DPN-GAN shows significant performance improvement, achieving PESQ, STOI and WARP-Q scores of $2.41$, $0.98$ and $0.892$, respectively. If we further increase the training data size to $95\%$ of the dataset, a similar performance is observed. This implies that the performance of the model saturates for training data over $90\%$ of the dataset size. Therefore, the optimal size for training the proposed DPN-GAN is $90\%$ of the dataset.

%%%%%%%% Table 1 %%%%%%%%%%%%%
\begin{table}
  \caption{Effect of training data size on performance of DPN-GAN}
\centering
  \label{tab1}
\begin{tabular}{@{}cccc@{}}
\toprule
Training Size(\%) & PESQ (\(\uparrow\)) & STOI (\(\uparrow\)) & WARP-Q (\(\downarrow\)) \\ \midrule
1                                 & 1.36                     & 0.85                     & 1.29                                                                   \\
10                                & 1.37                     & 0.89                     & 1.167                                                                  \\
20                                & 1.44                     & 0.91                     & 1.013                                                                  \\
50                                & 1.73                     & 0.94                     & 0.927                                                                  \\
75                                & 2.18                     & 0.96                     & 0.9                                                                    \\
90                                & 2.41                     & 0.98                     & 0.89                                                                   \\
95                                & 2.41                     & 0.98                     & 0.892                                                                  \\ \bottomrule
\end{tabular}
\end{table}
%%%%%%%%%%%%%%%%%%%%%%%%%%

Next, we consider the impact of the DPN layer depth on the performance of the proposed DPN-GAN. The DPN module is a critical component of the generator network, which process multi-receptive feature extraction from the input mel-spectrogram. The depth of the DPN layer ranging from $1$ to $5$ is considered for evaluation. As can be observed from Table \ref{tab2}, a shallow DPN module (depth of $1$ or $2$) results in underfitting of the actual audio distribution. The DPN depth of $4$ yields the optimum performance of the model on the testing set. Increasing the DPN depth further than $4$ either results in overfitting or saturation of model performance. Therefore, the DPN depth is set to $4$ for training purpose.

%%%%%%%% Table 2 %%%%%%%%%%%%%
\begin{table}
  \caption{Effect of DPN layer depth on performance of DPN-GAN}
\centering
  \label{tab2}
\begin{tabular}{@{}cccc@{}}
\toprule
DPN Depth & PESQ (\(\uparrow\)) & STOI (\(\uparrow\)) & WARP-Q (\(\downarrow\)) \\ \midrule
1                             & 1.36                     & 0.94                     & 0.943    \\
2                             & 1.51                     & 0.94                     & 0.927    \\
3                             & 1.89                     & 0.97                     & 0.91     \\
4                             & 2.41                     & 0.98                     & 0.892    \\
5                             & 2.42                     & 0.98                     & 0.884    \\ \bottomrule
\end{tabular}
\end{table}
%%%%%%%%%%%%%%%%%%%%%%%%%%

We also analyze the depth of the sub-discriminator for both the MSD and MCD. These two different modules of the discriminator process the audio data from two different aspects. Therefore, understanding the proper configuration of these layers are important. One assumption taken during this analysis is that the depth of both the sub-discriminators will be same in order to provide equal weightage to the feature extraction by the MSD and MCD, respectively. The experimental results with varied depths of the sub-discriminators are shown in Table \ref{tab3}. These results indicate that the performance of the proposed DPN-GAN significantly improves with the increasing depth of the sub-discriminators. Considering a depth of $1$, the achieved PESQ, STOI and WARP-Q scores of $1.37$, $0.95$ and $0.933$, respectively, shows that the model is underfitting with such a shallow discriminator architecture. The optimum performance of the model is achieved with the sub-discriminators depth of $3$, beyond which the performance of the model does not improve considerably and saturates.

%%%%%%%% Table 3 %%%%%%%%%%%%%
\begin{table}
  \caption{Effect of MSD and MCD depths on performance of DPN-GAN}
\centering
  \label{tab3}
\begin{tabular}{@{}cccc@{}}
\toprule
Depth & PESQ (\(\uparrow\)) & STOI (\(\uparrow\)) & WARP-Q (\(\downarrow\)) \\ \midrule
1                         & 1.37                     & 0.95                     & 0.933    			\\
2                         & 1.73                     & 0.96                     & 0.914   			 \\
3                         & 2.41                     & 0.98                     & 0.892   			 \\
4                         & 2.42                     & 0.98                     & 0.89    			 \\
5                         & 2.42                     & 0.98                     & 0.884   			 \\ \bottomrule
\end{tabular}
\end{table}
%%%%%%%%%%%%%%%%%%%%%%%%%%

To provide empirical evidence of the superiority of the proposed AdaPReLU activation function, we compare the performance of the proposed DPN-GAN using different activation functions. As we can observe from Table \ref{tab4}, the proposed AdaPReLU activation function demonstrates the best overall performance, achieving the highest PESQ score (2.41), STOI score (0.98), and the lowest WARP-Q score (0.892), indicating its superior ability to enhance audio signal clarity and intelligibility. Periodic ReLU also performs well, with a PESQ score of 2.09 and STOI score of 0.91, surpassing traditional activation functions like ReLU, Sigmoid, and Tanh. ReLU, a commonly used activation function, achieves a relatively high PESQ score of 1.75 and STOI score of 0.84, but underperforms compared to the periodic-based functions. SiLU, another modern activation function, exhibits moderate performance, with a PESQ score of 1.96 and STOI score of 0.76. The Sigmoid and Hyperbolic Tangent (Tanh) functions yield the lowest PESQ and STOI scores, indicating their limited effectiveness in this context.

%%%%%%%% Table 4 %%%%%%%%%%%%%
\begin{table}
  \caption{Effect of different activation functions on performance of DPN-GAN}
\centering
  \label{tab4}
\begin{tabular}{@{}lccc@{}}
\toprule
Activation Function	& PESQ (\(\uparrow\))		& STOI (\(\uparrow\))		& WARP-Q (\(\downarrow\))	\\ \midrule
Sigmoid			& 1.18                     		& 0.61                     		& 1.573   			\\
Hyperbolic Tangent	& 1.34                     		& 0.69                     		& 1.439  			 \\
ReLU				& 1.75                     		& 0.84                     		& 1.004   			 \\
SiLU				& 1.96                     		& 0.76                     		& 1.068    			 \\
Periodic ReLU		& 2.09                     		& 0.91                     		& 0.937  			 \\ 
Adaptive Periodic ReLU	& 2.41                     		& 0.98                     		& 0.892                      \\  \bottomrule
\end{tabular}
\end{table}
%%%%%%%%%%%%%%%%%%%%%%%%%%

Next, we compare the performance of the proposed DPN-GAN with state-of-the-art generative models. The results are demonstrated in Table \ref{tab5}. All the performance metrics are obtained from experiments on the testing set which is $0.09\%$ of the total dataset. Since all the models are very recent and considered state-of-the-art, they are performing considerably well on this dataset. 

From the Table \ref{tab5}, we can see that the proposed DPN-GAN outperforms other methods in terms of audio quality and speech intelligibility. HiFi-GAN, a baseline model in the comparison, achieves a PESQ, STOI and WARP-Q scores of $2.13$, $0.93$ and $1.233$, respectively. UnivNet trained with a learning rate of $1e^{-4}$ and FRE-GAN achieved relatively better performance. BigVGAN trained with a learning rate of $1e^{-4}$ shows an improved performance compared to UnivNet and HiFi-GAN. Since it is a large-scale model, the time taken for training is also higher than HiFi-GAN and UnivNet. SpecDiff-GAN performs significantly better than the above-mentioned models on all metrics, which shows the impact of diffusion process during training. The proposed DPN-GAN small model performs considerably well compared to other models on most metrics, whereas the DPN-GAN large outperforms all the state-of-the-art models by a large margin in terms of all the metrics. 

\begin{figure}[h]
\centering
\includegraphics[width=0.8\linewidth]{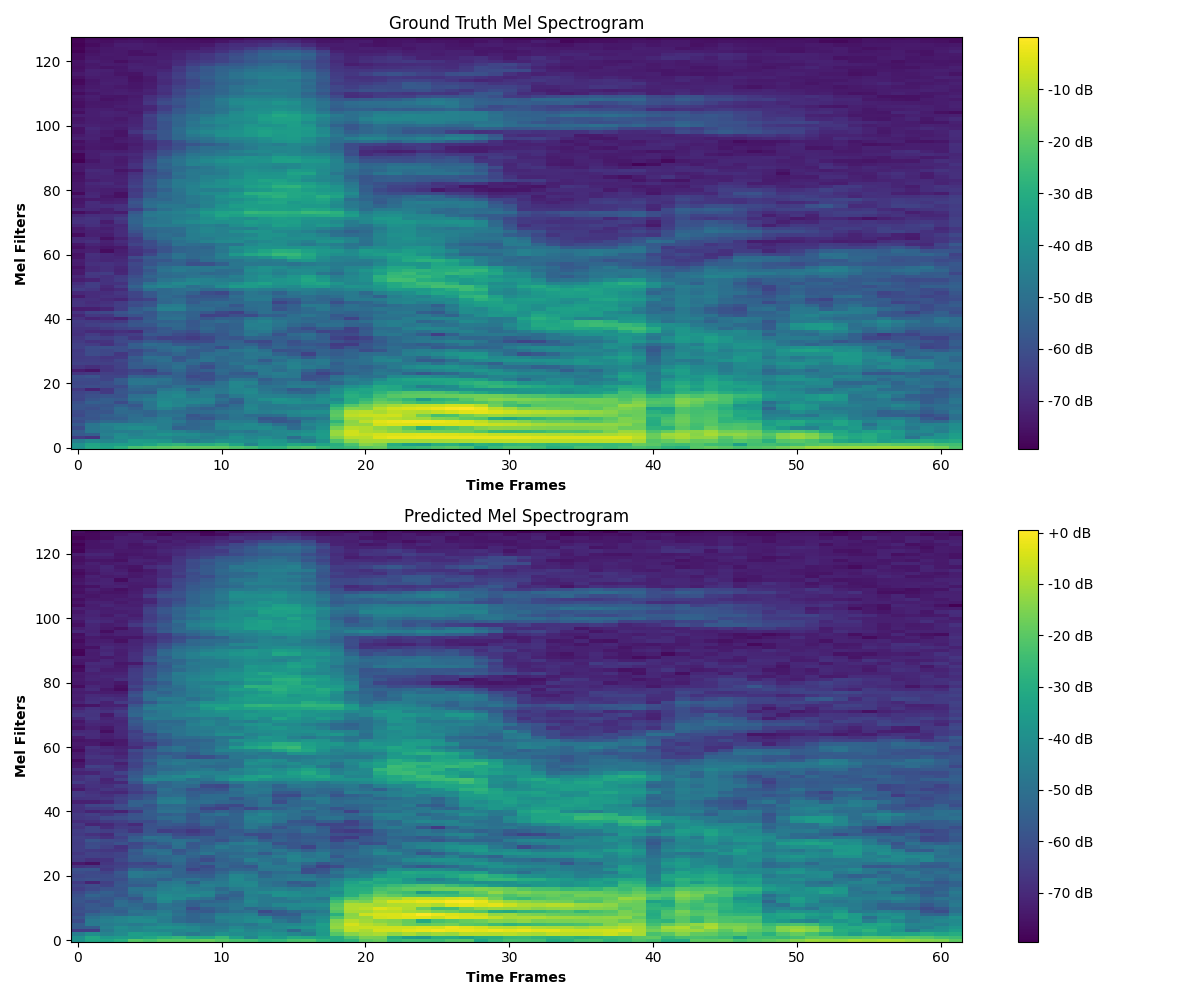}
\caption{Comparison of mel-spectrogram generated by DPN-GAN}
\label{fig4}
\end{figure}

In Fig. \ref{fig4}, we compare the mel-spectrogram generated by the proposed DPN-GAN with the ground truth mel-spectrogram. The mel-spectrogram generated by the DPN-GAN is more clear, and closer to the real mel-spectrogram.

%%%%%%%% Table 5 %%%%%%%%%%%%%
\begin{table}
  \caption{Performance comparison on AudioMNIST}
\centering
  \label{tab5}
\begin{tabular}{@{}lccc@{}}
\toprule
Model              			& PESQ (\(\uparrow\))        & STOI (\(\uparrow\))		& WARP-Q (\(\downarrow\)) \\ \midrule
HIFI-GAN           			& 2.13 				& 0.93                    			& 1.233                                     \\
UNIV-NET (lr=1e-4) 		& 2.42 				& 0.94                     		& 1.106                                       \\
SPECDIFF-GAN			& 2.66				& 0.97				& 0.931                                       \\
BIGV-GAN (lr=1e-4)		& 2.38				& 0.95                     		& 0.994                                       \\
FRE-GAN				& 2.19				& 0.95				& 0.985                                        \\
DPN-GAN small			& 2.41				& 0.98				& 0.892				 \\
DPN-GAN large			& 2.83				& 0.99				& 0.761				    \\ \bottomrule
\end{tabular}
\end{table}
%%%%%%%%%%%%%%%%%%%%%%%%%%

%%%%%%%% Figure 4 %%%%%%%%%%%%%
%%%%%%%%%%%%%%%%%%%%%%%%%%

%%%%%%%%% Subsubsection 6.1.2 %%%%%%%%%%%
\subsubsection{Results on LJSpeech} \label{subsubsec:6.1.2}

\begin{figure}[h]
\centering
\includegraphics[width=0.8\linewidth]{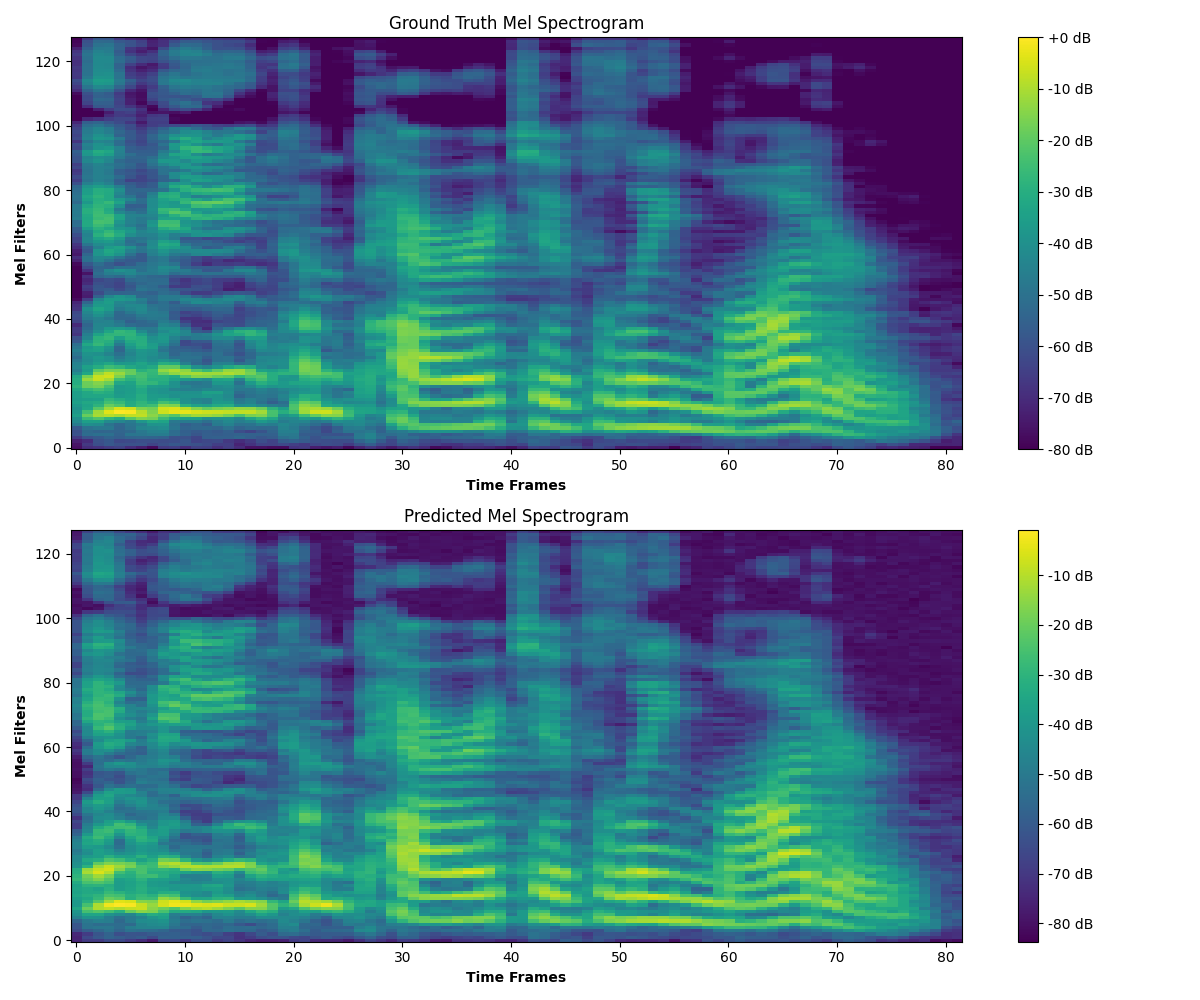}
\caption{Comparison of mel-spectrogram generated by DPN-GAN }
\label{fig5}
\end{figure}

For the LJSpeech dataset, we consider a training, validation and testing splits of $0.8$, $0.01$ and $0.19$, respectively. From the results shown in Table \ref{tab6}, we can see that both the DPN-GAN small and DPN-GAN large outperform all the benchmark and state-of-the-art models across all the metrics. SpecDiff-GAN shows a significant performance margin compared to other baseline models, HiFi-GAN, UNIV-NET, BigVGAN, and Fre-GAN. Moreover, FRE-GAN closely follows the SpecDiff-GAN with an insignificant difference on most metrics. The mel-spectrogram generated by the proposed DPN-GAN is compared with the ground truth in Fig. \ref{fig5}.

%%%%%%%% Table 6 %%%%%%%%%%%%%
\begin{table}
  \caption{Performance comparison on LJSpeech}
\centering
  \label{tab6}
\begin{tabular}{@{}lccc@{}}
\toprule
Model			& PESQ (\(\uparrow\)) 	& STOI (\(\uparrow\)) 	& WARP-Q (\(\downarrow\)) \\ \midrule
HIFI-GAN			& 3.47                           	& 0.98                           	& 1.203    				     \\
UNIV-NET (lr=1e-4)	& 3.44                           	& 0.98                           	& 1.33				     \\
SPECDIFF-GAN		& 3.76                           	& 0.99                           	& 1.018				    \\
BIGV-GAN (lr=1e-4)	& 3.72                           	& 0.98                           	& 1.073				    \\
FRE-GAN			& 3.68                           	& 0.98                           	& 1.157				    \\
DPN-GAN small        	& 3.79     			& 0.99			& 1.003				    \\
DPN-GAN large        	& 3.91	  		& 0.99			& 0.982				   \\ \bottomrule
\end{tabular}
\end{table}
%%%%%%%%%%%%%%%%%%%%%%%%%%

%%%%%%%% Figure 5 %%%%%%%%%%%%%
%%%%%%%%%%%%%%%%%%%%%%%%%%

%%%%%%%%% Subsubsection 6.1.3 %%%%%%%%%%%
\subsubsection{Results on LibriSpeech} \label{subsubsec:6.1.3}

We now evaluate the performance of the proposed DPN-GAN on the LibriSpeech dataset.  We split the dataset into a training ($0.85$), a validation ($0.01$), and a test ($0.14$) set. The learning rate was set to $1e^{-5}$ for both the generator optimizer and the discriminator optimizer while training the proposed DPN-GAN. After training for 500 epochs, an approximate minima was obtained for the model. The results are presented in Table \ref{tab7}. 

\begin{figure}[h]
\centering
\includegraphics[width=0.8\linewidth]{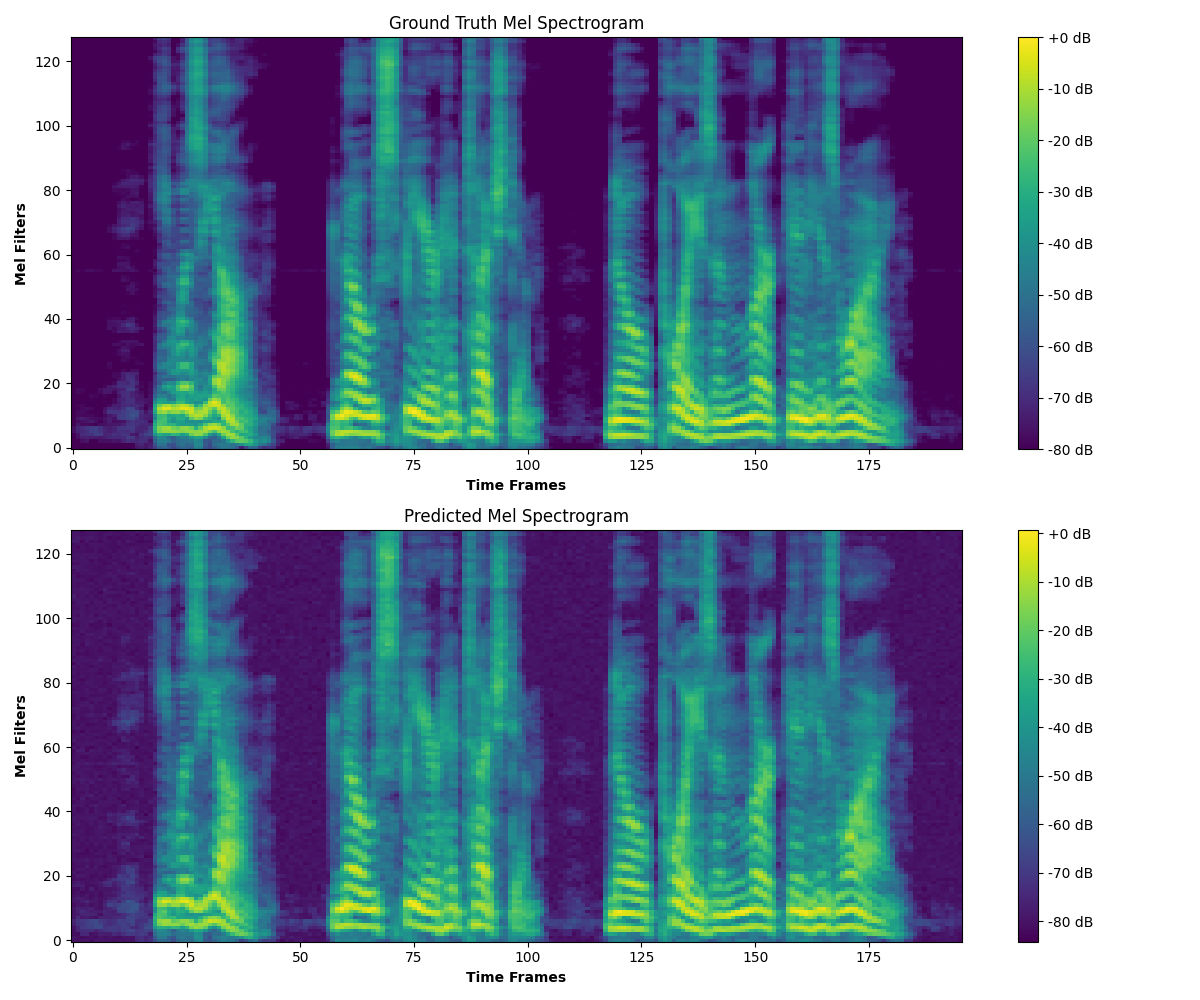}
\caption{Comparison of mel-spectrogram generated by DPN-GAN }
\label{fig6}
\end{figure}

It can be observed from Table \ref{tab7} that both versions of the proposed DPN-GAN consistently outperform other methods by a large margin on all the metrics. We also note that SpecDiff-GAN is also relatively stable on different datasets, performing well than other baseline models. Besides, the BigVGAN closely follows the SpecDiff-GAN. 
The mel-spectrogram generated by the proposed DPN-GAN closely resemble the ground truth mel-spectrogram, as shown in Fig. \ref{fig6}.

%%%%%%%% Table 7 %%%%%%%%%%%%%
\begin{table}
  \caption{Performance comparison on LibriSpeech}
\centering
  \label{tab7}
\begin{tabular}{@{}lccc@{}}
\toprule
Model			& PESQ (\(\uparrow\))	& STOI (\(\uparrow\)) 	& WARP-Q (\(\downarrow\))  \\ \midrule
HIFI-GAN			& 2.19				& 0.96                     		& 1.187  					\\
UNIV-NET (lr=1e-4)	& 2.47				& 0.96                     		& 1.24   					\\
SPECDIFF-GAN		& 3.27				& 0.98                     		& 1.013  					\\
BIGV-GAN (lr=1e-4)	& 3.03				& 0.97                     		& 1.008 					 \\
FRE-GAN			& 2.76				& 0.96                     		& 1.115 					 \\
DPN-GAN small		& 3.64				& 0.98                     		& 0.996					  \\ 
DPN-GAN large		& 3.88				& 0.99				& 0.947					  \\ \bottomrule
\end{tabular}
\end{table}
%%%%%%%%%%%%%%%%%%%%%%%%%%

%%%%%%%% Figure 6 %%%%%%%%%%%%%
%%%%%%%%%%%%%%%%%%%%%%%%%%

%%%%%%%%% Subsubsection 6.1.4 %%%%%%%%%%%
\subsubsection{Results on VCTK} \label{subsubsec:6.1.4}

Next, we train all the models on VCTK dataset. Since each datapoint in the dataset is large compared to the previous datasets, and each audio sequence is of considerable length, we introduce some extra depth to the DPN module and the sub-discriminators. Hence, the depth of DPN module is $5$, and that for each sub-discriminator is $4$. We also increase the size of hidden layers for additional complexity. The learning rate remains the same as that used for the LibriSpeech dataset, i.e., $1e^{-5}$, to train the proposed DPN-GAN. Table \ref{tab8} shows the results of the models' performance on VCTK dataset. 

\begin{figure}[h]
\centering
\includegraphics[width=0.8\linewidth]{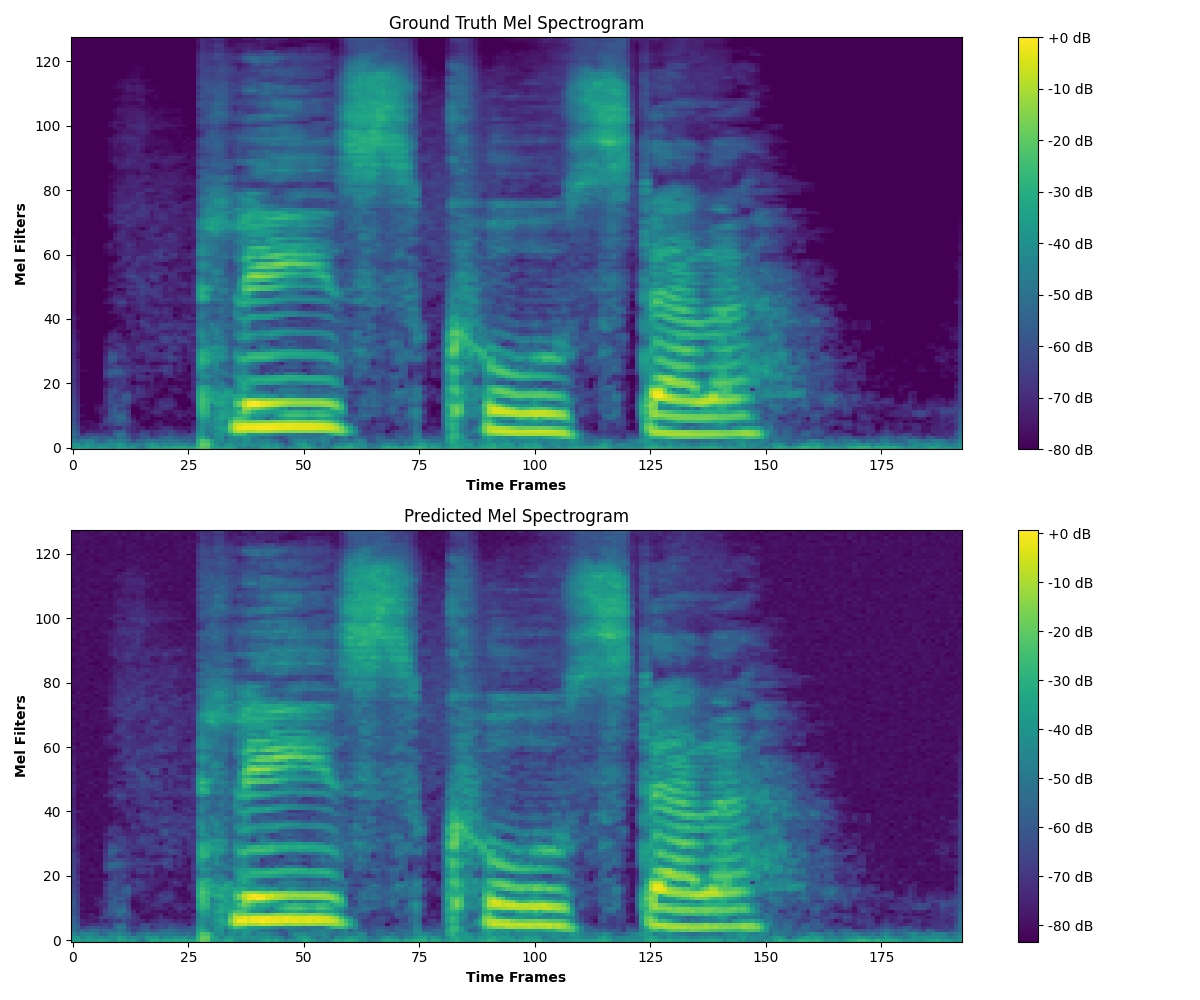}
\caption{Comparison of mel-spectrogram generated by DPN-GAN }
\label{fig7}
\end{figure}

From Table \ref{tab8}, it is observed that the DPN-GAN small lags behind BigVGAN on certain metrics, which is unlike the results on other datasets. This implies the benefit provided by diversity, quality, and suitability of a large-scale dataset for the BigVGAN model architecture. The SpecDiff-GAN closely follows the proposed DPN-GAN small. It is worth noting that DPN-GAN large model exceeds BigVGAN, achieving the highest performance on the VCTK dataset. Figure \ref{fig7} illustrates the mel-spectrogram generated by the proposed DPN-GAN.

%%%%%%%% Table 8 %%%%%%%%%%%%%
\begin{table}
  \caption{Performance comparison on VCTK}
\centering
  \label{tab8}
\begin{tabular}{@{}lccc@{}}
\toprule
Model			& PESQ (\(\uparrow\))	& STOI (\(\uparrow\))		& WARP-Q (\(\downarrow\))	\\ \midrule
HIFI-GAN			& 2.97				& 0.94                     		& 1.213					\\
UNIV-NET (lr=1e-4)	& 3.21                     		& 0.94                     		& 1.209					\\
SPECDIFF-GAN		& 3.52                     		& 0.96                     		& 0.983					\\
BIGV-GAN (lr=1e-4)	& 3.67                     		& 0.96                     		& 0.959					\\
FRE-GAN			& 3.49                     		& 0.94                     		& 1.157					\\
DPN-GAN	small		& 3.55                     		& 0.96                     		& 0.981					\\ 
DPN-GAN large        	& 3.71				& 0.98				& 0.915					\\ \bottomrule
\end{tabular}
\end{table}
%%%%%%%%%%%%%%%%%%%%%%%%%%

%%%%%%%% Figure 7 %%%%%%%%%%%%%
%%%%%%%%%%%%%%%%%%%%%%%%%%

%%%%%%%%% Subsubsection 6.1.5 %%%%%%%%%%%
\subsubsection{Results for GTZAN Dataset} \label{subsubsec:6.1.5}

Finally, the effectiveness of the proposed DPN-GAN is evaluated on a music generation dataset. We split the GTZAN dataset into $0.95$/$0.01$/$0.04$ for train/validation/test splits. The model was trained with a learning rate of $1e^{-4}$ and a batch size of $128$.  The depth of the DPN layer was kept at $4$, and that of the sub-discriminator layer at $3$. Since the task involves music generation, FAD and FDSD metrics were used to evaluate the model’s performance. The experimental results in terms of FAD and FDSD are listed in Table \ref{tab9}, with lower scores indicate higher audio/music quality.

\begin{figure}[h]
\centering
\includegraphics[width=0.8\linewidth]{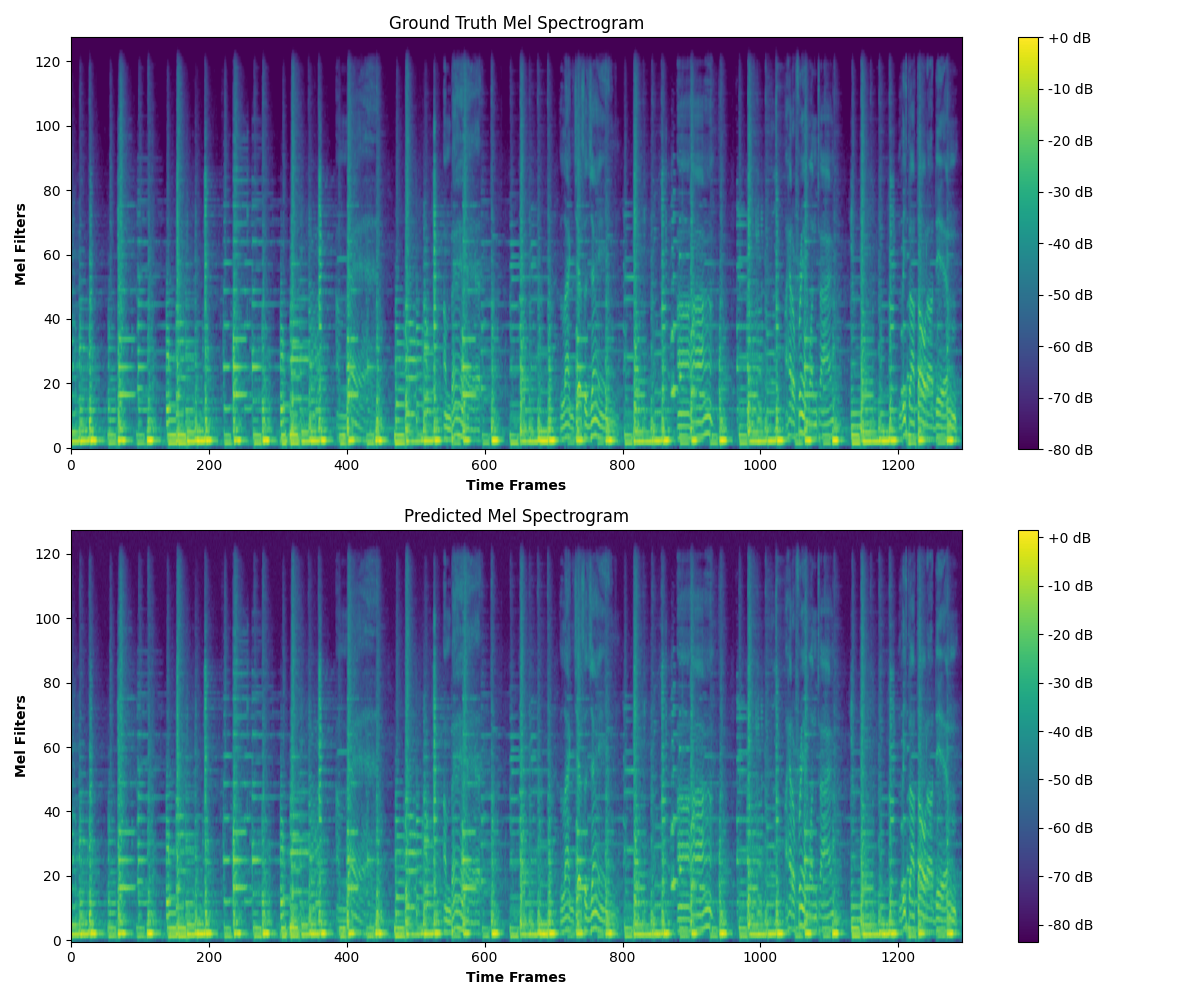}
\caption{Comparison of mel-spectrogram generated by DPN-GAN}
\label{fig8}
\end{figure}

It can be observed from Table \ref{tab9} that both versions of the DPN-GAN exhibit exceptional performance. The DPN-GAN large consistently outperforms the other compared models by a large margin across all the metrics. BigVGAN demonstrates superior performance compared to other baseline models. The FRE-GAN and SpecDiff-GAN closely follows BigVGAN. HiFi-GAN performs worse than UNIV-NET. As illustrated in Fig. \ref{fig8}, the mel-spectrogram predicted by the proposed DPN-GAN is more realistic and closer to the ground truth.

%%%%%%%% Figure 8 %%%%%%%%%%%%%
%%%%%%%%%%%%%%%%%%%%%%%%%%

%%%%%%%% Table 9 %%%%%%%%%%%%%
\begin{table}
  \caption{Performance comparison on GTZAN}
\centering
  \label{tab9}
\begin{tabular}{@{}lcc@{}}
\toprule
Model			& FAD (\(\downarrow\))	& FDSD (\(\downarrow\)) 	\\ \midrule
HIFI-GAN			& 0.143                   		& 0.197				   \\
UNIV-NET (lr=1e-4)	& 0.129                   		& 0.165				   \\
SPECDIFF-GAN		& 0.101                   		& 0.102				   \\
BIGV-GAN (lr=1e-4)	& 0.094                   		& 0.099				   \\
FRE-GAN			& 0.107                   		& 0.113				   \\
DPN-GAN	small		& 0.083                   		& 0.093				   \\ 
DPN-GAN large		& 0.046				& 0.079   				\\ \bottomrule
\end{tabular}
\end{table}
%%%%%%%%%%%%%%%%%%%%%%%%%%

%%%%%%%%%%%%%%%%%%%%%%%%%%%%%%%%%%
%%%%%%%%%%%%  Subsection 6.2 %%%%%%%%%%%%%%
%%%%%%%%%%%%%%%%%%%%%%%%%%%%%%%%%%
\subsection{Ablation Experiments} \label{subsec:6.2}

In this section, we conducted ablation experiments on the AudioMNIST dataset to understand the significance of each module and various loss functions of the proposed DPN-GAN. For ablation analysis, we use the DPN-GAN small model with a learning rate of $1e^{-5}$ for both the generator and discriminator optimizers. The instability of GANs training is mainly associated with the regular model architecture, such as rectangular sliding window convolution operation, and traditional cross-entropy and adversarial loss functions. We show that each loss component plays a significant role in the overall convergence and stability of the proposed DPN-GAN training. Moreover, we used deformable convolutions in our proposed model, which perform non-regular sliding window operation on the audio sequence. This is particularly useful when there is a variability in the dataset.

%%%%%%%%% Subsubsection 6.2.1 %%%%%%%%%%%
\subsubsection{Ablation on Model Architecture} \label{subsubsec:6.2.1}

The results for the ablated DPN-GAN small with key components removed from the full model are shown in Table \ref{tab10}. The baseline model without any component removed achieves a PESQ, STOI and WARP-Q scores of $2.41$, $0.98$ and $0.892$, respectively. By removing the DPN module or deformable convolution in MCD, the performance of the proposed DPN-GAN drops significantly. Using standard convolution operation instead of deformable convolution in MCD, we achieve a PESQ score of $1.71$, STOI score of $0.86$, and WARP-Q score of $0.729$, whereas we receive a PESQ score of $1.76$, STOI score of $0.91$, and WARP-Q score of $0.644$ by replacing the DPN module with convolutional layers. These results validate the importance of DPN module and deformable convolution for the generation of high-fidelity audio sequences. Removing the PRAK module also result in significant performance degradation. This verifies the fact that use of periodic activation functions significantly improves the performance of the model. In addition, removing MCD from the discriminator network causes approximately equal performance drop as the case of removing the DPN module. The MSD module in the discriminator network illustrate the importance of sequential operation during classification. Although it has less impact on the performance of the proposed DPN-GAN as compared to the MCD module, the results obtained show a lack of continuity in the generated audio sequence. Moreover, ablation of the metadata channel also slightly worse in all metrics compared to the baseline model. As such, removing the metadata from the analysis reduce the capability of the model for conditional generation. These results indicate that each of the key components in the proposed DPN-GAN is necessary to achieve high quality and intelligible speech synthesis. 

%%%%%%%% Table 10 %%%%%%%%%%%%%
\begin{table*}
  \caption{Results of ablation study on key components of DPN-GAN}
\centering
  \label{tab10}
\begin{tabular}{@{}lccc@{}}
\toprule
Case								& PESQ (\(\uparrow\))		& STOI (\(\uparrow\))		& WARP-Q (\(\downarrow\)) \\ \midrule
Without Removing anything				& 2.41				& 0.98				& 0.892  					\\
Without MetaData					& 2.15				& 0.96				& 0.812					  \\
Without DPN Module					& 1.76				& 0.91 				& 0.644					 \\
Using ReLU instead of PRAK				& 1.86				& 0.91				& 0.705					  \\
Without Deformable Convolution in MCD		& 1.71				& 0.86				& 0.729					  \\
Removing MSD from Discriminator			& 2.04				& 0.89				& 0.753					  \\
Removing MCD from Discriminator			& 1.93				& 0.87				& 0.694					  \\ \bottomrule
\end{tabular}
\end{table*}
%%%%%%%%%%%%%%%%%%%%%%%%%%

%%%%%%%%% Subsubsection 6.2.2 %%%%%%%%%%%
\subsubsection{Ablation on Loss Functions} \label{subsubsec:6.2.2}

In addition to illustrating the importance of key architectural components, we also highlight the importance of the loss functions used to train the proposed DPN-GAN. We used several loss components in both the generator and discriminator networks. In the generator network, we have three different loss components, an adversarial loss, mel-spectrogram loss and a feature matching loss. 

The results of ablation experiments for various loss functions are shown in Table \ref{tab11}. First, removing the mel-spectrogram loss component from the loss function substantially reduces the model performance, as inferred from the metric values in Table \ref{tab11}. Moreover, we observed an increase in the feature matching loss value which is $0.44$, whereas the feature matching loss value for the baseline model is $0.17$. Consequently, this increases the generator loss value, and the output of the model becomes distorted in nature.  Next, we removed the feature matching loss component from the generator loss function. The PESQ, STOI and WARP-Q values obtained for this scenario is $1.94$, $0.92$, $0.764$, respectively, as shown in Table \ref{tab11}. From the results, we can infer that removing the feature matching loss reduces the human audibility of the generated speech, since the PESQ value reduces significantly from the baseline model. Moreover, we also observed that the mel-spectrogram loss increases significantly, and becomes $0.69$ (baseline mel-spectrogram loss is $0.24$). Hence, both the loss components are significant for the model performance.

%%%%%%%% Table 11 %%%%%%%%%%%%%
\begin{table}
  \caption{Results of ablation study on various loss components of DPN-GAN}
\centering
  \label{tab11}
\begin{tabular}{@{}lccc@{}}
\toprule
Case                          				& PESQ (\(\uparrow\)) & STOI (\(\uparrow\)) 	& WARP-Q (\(\downarrow\)) \\ \midrule
Without Mel-Spectrogram Loss  		& 1.83                           & 0.91                           	& 0.708   \\
Without Feature Matching Loss 		& 1.94                           & 0.92                           	& 0.764   \\ 
\bottomrule
\end{tabular}
\end{table}
%%%%%%%%%%%%%%%%%%%%%%%%%%

%%%%%%%%%%%%%%%%%%%%%%%%%%%%%%%%%%
%%%%%%%%%%%%  Subsection 6.3 %%%%%%%%%%%%%%
%%%%%%%%%%%%%%%%%%%%%%%%%%%%%%%%%%
\subsection{Robustness to Out-of-Distribution and Noisy Data} \label{subsec:6.3}

This section evaluates the robustness of the proposed DPN-GAN small to out-of-distribution, and noisy data. We compare the model’s performance for unseen languages or recording environments, and in the presence of unclean or noisy data on the LJSpeech dataset.

%%%%%%%%% Subsubsection 6.3.1 %%%%%%%%%%%
\subsubsection{Performance Comparison on Out-of-Distribution Data} \label{subsubsec:6.3.1}

Beside the performance comparison on various datasets and an ablation study, we evaluate the performance of our proposed DPN-GAN small on unseen languages and varied recording environments. In the LJSpeech dataset used for this analysis, there are speakers from other languages like Spanish, mandarin, etc. We exclude speakers belonging to these language categories from the training data, and created a separate dataset. For training the proposed DPN-GAN small, we use the batch size of $128$, and a learning rate of $1e^{-5}$ for both the generator and discriminator optimizers. Additionally, we set the depth of the DPN module and those of sub-discriminators to $4$ and $3$, respectively. PESQ and WARP-Q metrics are considered to evaluate the model’s performance. 

The experimental results of various models on out-of-distribution scenario are listed in Table \ref{tab12}. It is observed from Table \ref{tab12} that HiFi-GAN lacks the robustness to unseen data. It achieves the lowest PESQ and highest WARP-Q scores of $1.74$ and $1.054$, respectively, indicating its highly data-driven nature. UnivNet model and FRE-GAN with a slightly improved performance than the HiFi-GAN also lacks robustness to out-of-distribution data. In contrast to the performance on the original LJSpeech dataset, BigVGan trained with a learning rate of $1e^{-4}$ performs better than the SpecDiff-GAN and other compared state-of-the-art baseline models. The proposed DPN-GAN small outperforms BigVGAN by a clear margin, achieving a PESQ and WARP-Q scores of $2.11$ and $0.917$, respectively. Hence, the proposed DPN-GAN can generate more audible speech signals compared to other state-of-the-art speech synthesis models, when the mel-spectrogram and metadata of an unseen language and recording environment is provided to the model.

%%%%%%%% Table 12 %%%%%%%%%%%%%
\begin{table}
  \caption{Robustness analysis on out-of-distribution data}
\centering
  \label{tab12}
\begin{tabular}{@{}lcc@{}}
\toprule
Model              		& PESQ (\(\uparrow\)) 	& WARP-Q (\(\downarrow\)) \\ \midrule
HIFI-GAN           		& 1.51                           	& 1.104   \\
UNIV-NET (lr=1e-4) 	& 1.83                           	& 1.083   \\
SPECDIFF-GAN       	& 2.26                           	& 0.981   \\
BIGV-GAN (lr=1e-4) 	& 2.14                           	& 1.075   \\
FRE-GAN            		& 1.67                           	& 1.149   \\ 
DPN-GAN small           	& 2.26                           	& 0.928   \\ 
\bottomrule
\end{tabular}
\end{table}
%%%%%%%%%%%%%%%%%%%%%%%%%%

%%%%%%%%% Subsubsection 5.3.2 %%%%%%%%%%%
\subsubsection{Performance Comparison on Noisy and Perturbed Data} \label{subsubsec:6.3.2}

To further evaluate the robustness of the proposed DPN-GAN to noisy and perturbed data, we add a zero-mean Gaussian noise with standard deviation of $1$ scaled to a factor of $0.05$ to the LJSpeech dataset. The training configuration is identical to that in the previous experiment. The experimental results of various models on noisy data are shown in Table \ref{tab13}.

From Table \ref{tab13}, it is observed that BigV-GAN model with a learning rate of $1e^{-4}$ is more robust to noisy dataset than other baseline models, and achieves a PESQ and WARP-Q values of $2.09$ and $0.947$, respectively. Other baseline models cause significant performance degradation on noisy data. The proposed DPN-GAN small model outperforms all the compared models, and achieves a PESQ and WARP-Q scores of $2.11$ and $0.917$, respectively, indicating greater robustness to noisy or perturbed input data.

%%%%%%%% Table 13 %%%%%%%%%%%%%
\begin{table}
  \caption{Robustness analysis on perturbed data}
\centering
  \label{tab13}
\begin{tabular}{@{}lcc@{}}
\toprule
Model              		& PESQ (\(\uparrow\)) 	& WARP-Q (\(\downarrow\)) \\ \midrule
HIFI-GAN           		& 1.74                           	& 1.054   \\
UNIV-NET (lr=1e-4) 	& 1.88                           	& 1.177   \\
SPECDIFF-GAN       	& 1.91                           	& 0.906   \\
BIGV-GAN (lr=1e-4) 	& 2.09                           	& 0.947   \\
FRE-GAN            		& 1.83                           	& 1.032   \\
DPN-GAN small           	& 2.11                           	& 0.917   \\ 
\bottomrule
\end{tabular}
\end{table}
%%%%%%%%%%%%%%%%%%%%%%%%%%

We further evaluate the performance of our proposed DPN-GAN small in the presence of different scales of noise. From the results shown in Table \ref{tab14}, it is observed that the model provides optimum performance with a standard amount of noise in the dataset like 0.1 to 0.2. The performance of the model is not reducing by a large margin when the intensity of noise is around 0.4, and still performing better than some of the state-of-the-art models on unperturbed data. As we further increase the noise scale, the model performance reduces. Although the model’s performance decreases in the presence of extremely degraded data, we can clean the noisy data to a certain degree using various filtering techniques, which can then be fed to the model for optimal performance.

%%%%%%%% Table 14 %%%%%%%%%%%%%
\begin{table}
  \caption{Effect of Perturbation on Performance of DPN-GAN}
\centering
  \label{tab14}
\begin{tabular}{@{}cccc@{}}
\toprule
Noise Intensity              	& PESQ (\(\uparrow\)) 	& STOI (\(\uparrow\))	& WARP-Q (\(\downarrow\)) \\ \midrule
0.1           			& 2.06                           	& 0.96			& 0.901   \\
0.2 				& 2.00                           	& 0.84			& 0.976   \\
0.4       			& 1.68                           	& 0.81			& 1.035   \\
0.6 				& 1.21                           	& 0.63			& 1.119   \\
0.9            			& 0.84                           	& 0.59			& 1.386   \\
\bottomrule
\end{tabular}
\end{table}
%%%%%%%%%%%%%%%%%%%%%%%%%%

Hence, this analysis validates the generalization ability of the proposed DPN-GAN to out of distribution data, and improved robustness against noise perturbations.

 %%%%%%%%%%%%%%%%%%%%%%%%%%%%%%%%%%
%%%%%%%%%%%%  Subsection 6.4 %%%%%%%%%%%%%%
%%%%%%%%%%%%%%%%%%%%%%%%%%%%%%%%%%
\subsection{Runtime Comparison} \label{subsec:6.4}

Finally, we compare the runtime of our proposed DPN-GAN with other baseline models on the LibriSpeech dataset. All the training and system configurations has been mentioned in section \ref{sec:5}. The evaluation results for generating 24 KHz audio are demonstrated in Table \ref{tab15}. Since we are training the models on a GPU based system, Table \ref{tab15} provides two key information corresponding to each model: one is the synthesis speed of the model with respect to real-time, the other is the number of parameters which shows the size and complexity of the model.

%%%%%%%% Table 15 %%%%%%%%%%%%%
\begin{table}
  \caption{Runtime comparison}
\centering
  \label{tab15}
\begin{tabular}{@{}lcc@{}}
\toprule
Model			& Synthesis Speed		 & Number of Parameters (M) \\ \midrule
HiFI-GAN V1           	& $\times$135.18               	& 14.01    \\
UNIV-NET c-32 		& $\times$206.41               	& 14.86    \\
BIGV-GAN base 		& $\times$70.27                	& 14.01      \\
DPN-GAN large      	& $\times$26.88                	& 124.31   \\
DPN-GAN small      	& $\times$83.24               	& 38.67    \\
\bottomrule
\end{tabular}
\end{table}
%%%%%%%%%%%%%%%%%%%%%%%%%%

From Table \ref{tab15}, it is observed that HiFi-GAN V1 shows an impressive result with a generation speed that is $135.18$ times faster than real-time and a relatively low parameter count of $14.01$M. Meanwhile, BigV-GAN base with a similar number of parameters ($14.01$M) is notably slower than HiFi-GAN V1, synthesizing the audio at a speed that is only $70.27$ times faster than real-time. UNIV-NET c-32 with $14.86$M parameters demonstrates the highest generation speed of 206.41 time faster than real-time, outperforming other compared models in terms of raw generation efficiency. Moreover, DPN-GAN small with 38.67M parameters achieves a reasonable generation speed of 83.24 times faster than real-time. Despite the larger parameter count, DPN-GAN small outperforms the BigV-GAN in terms of synthesis speed. On the other hand, DPN-GAN large with significantly larger parameter size of 124.31M consistently demonstrates superior performance across various generation tasks, but the synthesis speed is the slowest among other compared models. This limitation in the proposed DPN-GAN mainly comes from using the PRAK as an activation kernel. Because it is a complex non-linear function with additional parameters required for its training, which substantially increases the complexity of the model. Moreover, we can observe that there is a tradeoff between computational fidelity of the model and the runtime. If we need a very high-fidelity model, the computation complexity of the model increases significantly. DPN-GAN large model has 124 M parameters which is too computationally expensive to train on a standard CPU system. Hence, additional hardware systems like GPU or TPU is required to train the model. Furthermore, we provided a lightweight DPN-GAN small model, which is less computationally expensive than DPN-GAN large, but the fidelity decreases. So, DPN-GAN small is more suitable for real-time applications, whereas we should use DPN-GAN large for offline training to get the high-fidelity model.

This comparison highlights the trade-off between model size and speed, with smaller models generally being faster, though certain architectures like UNIV-NET c-32 optimize this balance more effectively.

%%%%%%%%%%%%%%%%%%%%%%%%%%%%%%%%%%%%%%%%%%%%%%%%%%%%%%%%%%%%%%%%%%%%%%%
%%%%%%%%%%%%%%%%%%%%%%%%%%%%%%  SECTION 7 %%%%%%%%%%%%%%%%%%%%%%%%%%%%%%%%%%
%%%%%%%%%%%%%%%%%%%%%%%%%%%%%%%%%%%%%%%%%%%%%%%%%%%%%%%%%%%%%%%%%%%%%%%
\section{Conclusion}\label{sec:7}
In this paper, we proposed a deformable periodic network-based GAN model to generate high-fidelity diverse audio samples, called DPN-GAN. Existing GANs based speech synthesis models often encounter certain issues, such as limited model output scalability, generalization beyond audio speech, and mode collapse. Specifically, we leveraged deformable convolutions, and introduced a DPN module in the generator network which process the features extracted from the mel-spectrogram at multiple resolution and receptive fields. Additionally, our proposed Adaptive PRAK kernel activation function induces spectral bias to the input vectors which are of periodic nature. On the discriminator end, we introduced the DefMSD and DefMCD, each of which consist of several sub-discriminators, evaluating audio samples at different resolutions and processing the periodic samples inside the audio signal, respectively, which help to better capture the periodic patterns. In this way, the proposed DPN-GAN is able to generate high-fidelity and diverse audio samples, including speech and music, as demonstrated by the experimental results on various datasets. Moreover, the proposed DPN-GAN generalize well to unseen languages and recording environments, outperforming existing state-of-the-art models for both in-distribution and out-of-distribution samples. Besides, it also demonstrated high robustness to noisy and perturbed data. However, the proposed DPN-GAN lags behind the other compared models in runtime. Due to large size of the model, the time consumed per training and to generate audio samples is higher compared to other light-weight models. In future, it would be interesting to explore suitable activation kernels with similar performance output to increase the computational speed of the proposed DPN-GAN. Being a large model with huge parameter space, the proposed DPN-GAN is more suitable for large scale industrial applications requiring high-quality speeches, such as music in telecommunication devices. 

\bibliographystyle{unsrt}
\bibliography{references.bib}

\begin{IEEEbiography}[{\includegraphics[width=1in,height=1.25in,clip,keepaspectratio]{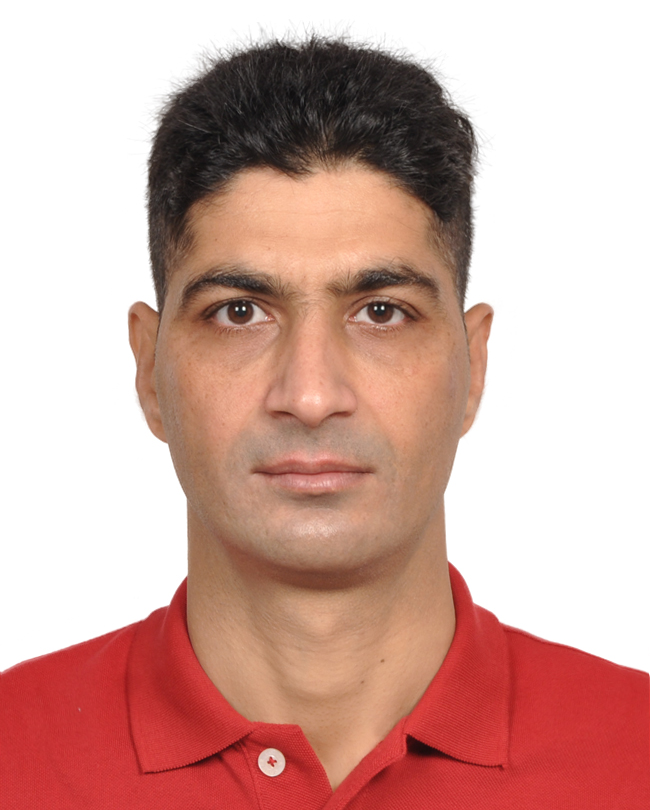}}]{Zeeshan Ahmad} (M’22) received the Ph.D. degree in information and communication engineering from Nanjing University of Science and Technology, Nanjing, China, in 2018. He was a Postdoctoral Researcher with the College of Information Science and Electronic Engineering, Zhejiang University, Hangzhou, China, from December 2018 to November 2020. He was a lecturer with the School of Cyber Science and Engineering, Ningbo University of Technology, Ningbo, China, from January 2021 to December 2024. He is currently an Associate Researcher with the Ningbo Institute of Digital Twin, Eastern Institute of Technology, Ningbo, China. His research interests include deep learning, computer vision, natural language processing, pattern recognition, generative AI, generative adversarial networks, wireless communications, and array signal processing. He has been a member of TPC of multiple international conferences including IEEE SAM. He is also a member of Chinese Institute of Electronics (CIE), China Computer Federation (CCF), and China Society of Image and Graphics (CSIG). He is currently serving on the Topical Advisory Panel for \textit{Computation}, MDPI, and as an Academic Editor for \textit{PLOS ONE}.

\end{IEEEbiography}
\begin{IEEEbiography}[{\includegraphics[width=1in,height=1.25in,clip,keepaspectratio]{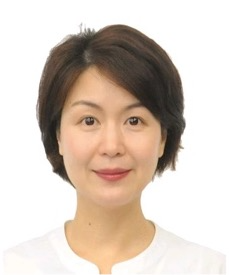}}]{Shudi Bao} (S’03--M’08) received the Ph.D. degree in communications and information systems from Southeast University, Nanjing, China, in 2007. After her Ph.D., she held positions with the Shenzhen Institute of Advanced Technology, Chinese Academy of Sciences, Beijing, China, and Agilent Technology, Singapore. From September 2011 to August 2024, she was with the Ningbo University of Technology, Ningbo, China, where she was the Dean of the School of Cyber Science and Engineering. She is currently an Associate Director at the Ningbo Institute of Digital Twin, Eastern Institute of Technology, Ningbo, China. Her expertise includes computational intelligence, bioinformatics, and information security.
\end{IEEEbiography}
\begin{IEEEbiography}[{\includegraphics[width=1in,height=1.25in,clip,keepaspectratio]{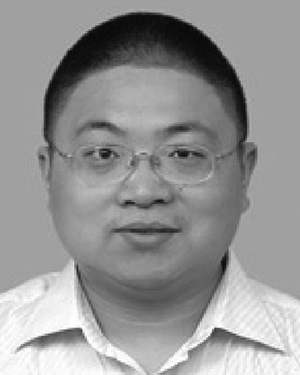}}]{Meng Chen} received the Master’s degree in computer application engineering from Zhejiang University of Technology, Hangzhou, China. He is currently an Associate Professor with the School of Cyber Science and Engineering (School of Computer Science and Engineering), Ningbo University of Technology, Ningbo, China. His research interests include mobile health system security, biometrics, and information security. 
\end{IEEEbiography}

\EOD

\end{document}